\documentclass[acmsmall,screen]{acmart}

\usepackage{tabularx}  
\AtBeginDocument{%
  }

\usepackage{etoolbox}
\AtBeginEnvironment{minted}{\footnotesize} 
\setcopyright{acmlicensed}
\copyrightyear{2018}
\acmYear{2018}
\acmDOI{XXXXXXX.XXXXXXX}

\acmJournal{JACM}
\acmVolume{37}
\acmNumber{4}
\acmArticle{111}
\acmMonth{8}

\usepackage{graphicx}
\usepackage{subfigure}

\usepackage{amsmath}
\usepackage{amsthm}
 
\usepackage{ragged2e}
\usepackage{wasysym}
\usepackage{tabularx}

\usepackage{float}
\usepackage{multirow}
\usepackage{mathrsfs}
\usepackage{algorithmic}
\usepackage[ruled,linesnumbered]{algorithm2e}
\SetKwRepeat{Do}{do}{while}

\usepackage{enumitem}

\usepackage{rotating}
\usepackage{balance}
\usepackage{threeparttable}
\usepackage{makecell}
\usepackage{pdflscape}

\usepackage{lineno}

\usepackage{graphicx}

\usepackage[export]{adjustbox}

\usepackage{stfloats}
\usepackage{verbatim}
\usepackage{setspace}
\usepackage{threeparttable}
\usepackage{supertabular,booktabs}
\usepackage{rotating}
\usepackage{longtable}
\usepackage{supertabular}
\usepackage{pifont}
\usepackage{xcolor,pict2e}

\usepackage{lettrine}
\usepackage{hyperref}
\hypersetup{
    colorlinks=true,
    linkcolor=blue,    
    citecolor=blue,    
    urlcolor=black      
}
\makeatletter
\renewcommand\@cite[2]{%
    {\color{blue}[#1]}
}
\makeatother
\usepackage[hyphenbreaks]{breakurl}
\usepackage{fancyhdr}

\allowdisplaybreaks[4] 

\usepackage{tcolorbox}
\tcbuselibrary{breakable} 
\usepackage{xcolor}
\usepackage{soul}   

\definecolor{lightpink}{rgb}{1.0, 0.87, 0.87}
\definecolor{lightpurple}{rgb}{0.94, 0.85, 0.94}
\definecolor{lightgreen}{rgb}{0.88, 1.0, 0.88}
\definecolor{lightyellow}{rgb}{1.0, 1.0, 0.88}
\definecolor{lightblue}{rgb}{0.88, 0.94, 1.0}
\definecolor{lgray}{rgb}{0.95,0.95,0.95}
\usepackage{listings}
\begin{document}
\title{Large Language Models for Automated Web-Form-Test Generation: An Empirical Study}

\author{Tao Li}
\email{3220007015@student.must.edu.mo}
\orcid{0009-0001-7413-9692}
\affiliation{
  \institution{School of Computer Science and Engineering, Macau University of Science and Technology}
  \city{Taipa}
  \state{Macau}
  \country{China}
  \postcode{999078}
}

\author{Chenhui Cui}
\email{3230002105@student.must.edu.mo}
\orcid{0009-0004-8746-316X}
\affiliation{
  \institution{School of Computer Science and Engineering, Macau University of Science and Technology}
  \city{Taipa}
  \state{Macau}
  \country{China}
  \postcode{999078}
}

\author{Rubing Huang}
\email{rbhuang@must.edu.mo}
\orcid{0000-0002-1769-6126}
\affiliation{
  \institution{School of Computer Science and Engineering, Macau University of Science and Technology}
  \city{Taipa}
  \state{Macau}
  \country{China}
  \postcode{999078}
}
\affiliation{
  \institution{Macau University of Science and Technology Zhuhai MUST Science and Technology Research Institute}
  \city{Zhuhai}
  \state{Guangdong Province}
  \country{China}
  \postcode{519099}
}

\author{Dave Towey}
\email{dave.towey@nottingham.edu.cn}
\orcid{0000-0003-0877-4353}
\affiliation{
  \institution{School of Computer Science, University of Nottingham Ningbo China}
  \city{Ningbo}
  \state{Zhejiang}
  \country{China}
  \postcode{315100}
}

\author{Lei Ma}
\email{ma.lei@acm.org}
\orcid{0000-0002-8621-2420}
\affiliation{
  \institution{School of Computer Science, University of Tokyo}
  \city{Tokyo}
  \country{Japan}
  \postcode{113-0033}
}
\affiliation{
  \institution{Department of Electrical and Computer Engineering, University of Alberta}
  \city{Edmonton}
  \country{Canada}
  \postcode{T6G 2R3}
}

\begin{abstract}
Testing web forms is an essential activity for ensuring the quality of web applications.
It typically involves evaluating the interactions between users and forms.
Automated test-case generation remains a challenge for web-form testing:
Due to the complex, multi-level structure of web pages, it can be difficult to automatically capture their inherent contextual information for inclusion in the tests.
\textit{Large Language Models} (LLMs) have shown great potential for contextual text generation.
This motivated us to explore how they could generate automated tests for web forms, making use of the contextual information within form elements.
To the best of our knowledge, no comparative study examining different LLMs has yet been reported for web-form-test generation.
To address this gap in the literature, we conducted a comprehensive empirical study investigating the effectiveness of 11 LLMs on 146 web forms from 30 open-source Java web applications. 
In addition, we propose three HTML-structure-pruning methods to extract key contextual information.
The experimental results show that different LLMs can achieve different testing effectiveness, with the GPT-4, GLM-4, and Baichuan2 LLMs generating the best web-form tests. 
Compared with GPT-4, the other LLMs had difficulty generating appropriate tests for the web forms:
Their \textit{successfully-submitted rates} (SSRs)
---
the proportions of the LLMs-generated web-form tests that could be successfully inserted into the web forms and submitted
---
decreased by 9.10\% to 74.15\%.
Our findings also show that, for all LLMs, when the designed prompts include complete and clear contextual information about the web forms, more effective web-form tests were generated. 
Specifically, when using Parser-Processed HTML for Task Prompt (PH-P), the SSR averaged 70.63\%, higher than the 60.21\% for Raw HTML for Task Prompt (RH-P) and 50.27\% for LLM-Processed HTML for Task Prompt (LH-P). 
With RH-P, GPT-4's SSR was 98.86\%, outperforming models like LLaMa2 (7B) with 34.47\% and GLM-4V with 0\%. 
Similarly, with PH-P, GPT-4 reached an SSR of 99.54\%, the highest among all models and prompt types.
Finally, this paper also highlights strategies for selecting LLMs based on performance metrics, and for optimizing the prompt design to improve the quality of the web-form tests.
\end{abstract}



\received{21 September 2024}
\received[revised]{31 March 2025}
\received[accepted]{6 May 2025}

\maketitle

\section{Introduction
\label{SEC:Introduction}}

In the current rapidly-evolving digital era, web applications have become a cornerstone of daily interactions. 
By March 2024, the Internet Archive had archived more than 866 billion web pages~\cite{InternetArchive2024}. 
A web application consists of various \textit{HyperText Markup Language} (HTML) elements, such as \textit{links}, \textit{buttons}, and \textit{sliders}. 
A key interface element of these applications is the web form, which is structured using the \textit{Document Object Model} (DOM)~\cite{marini2002document}. 
Web forms not only serve as an interaction bridge between users and web applications~\cite{kahn1996structured,furche2013ontological}, but also play an important role in improving the user experience and data-collection efficiency (including for user registration, login, and payments) \cite{jarrett2009forms,seckler2014designing,jesse2011elements}.
The correctness of the web-form inputs significantly impacts the subsequent execution process of web applications~\cite{miao2007model}:
If the web-form inputs meet the contextual requirements and can be successfully submitted, some later operations 
---
such as data validation and database updates
---
can also be processed correctly.
In contrast, if the submission is not successful, then the subsequent operations may not be executed, and the application's integrity cannot be assured.
Therefore, the generation of effective web-form inputs before exploring the subsequent operations of the web application is essential for testing web applications.

Web-form testing has been widely used to ensure the quality of web forms~\cite{santiago2019machine,alian2024bridging}. 
It aims at simulating user inputs and evaluating user interactions~\cite{bargas2011working}.
In addition to functionality validation, web-form testing plays an important role in ensuring security.
Improperly validated web-form inputs can introduce vulnerabilities that could be exploited through attacks such as SQL injections, cross-site scripting (XSS), and cross-site request forgery (CSRF) \cite{cheah2021review}.
This could compromise user data, disrupt application integrity, and expose web systems to malicious attacks~\cite{sarmah2018survey,fredj2021owasp}. 
To mitigate such risks, rigorous testing is required to verify input-handling mechanisms, and to ensure compliance with best practices~\cite{deepa2016securing}.
However, there may be challenges to automatically generating web-form tests, due to the properties of web forms:
(1) Web forms generally have complex structures with various components~\cite{rothermel1998you,lukanov2016using,cruz2017enabling}.
The basic web-form components include \textit{tags}, \textit{elements}, \textit{attributes}, and \textit{placeholders}.
In addition, developers may also introduce customized components (such as control-logic code or dynamic behaviors) that could increase the complexity of the web-form structures. 
For instance, some forms include fields with conditional logic that only become visible when certain selections are made.
(2) Web forms also provide diverse contextual information for user interaction~\cite{furche2013ontological}.
In this paper, ``contextual information'' refers to the structure and attributes of web-form elements
---
derived from web-form HTML
---
such as their tags, types, and attributes (e.g., name, id), as well as the relationships among these elements.
For example, web forms can offer a drop-down list or a set of radio buttons for users to make a single selection from multiple choices. 
The interrelationships among these elements, where choices in one field can affect the availability or state of others, further complicate test generation.

Web-form testing is critical for ensuring the accurate interaction and evaluation of contextual information in complex web structures.
Although some GUI-testing techniques can help to validate the user-interface components of web forms
---
such as those for cross-browser consistency~\cite{xu2018cross}, or scriptless automation~\cite{moura2023cytestion}
---
they are often not sufficient for addressing the interdependencies among web-form elements:
They may fail to capture the intricate relationships among form elements, leaving them unable to explore the structural complexity and contextual dependencies inherent in web forms.
This highlights the necessity for specialized testing techniques that consider both the structural and contextual dimensions of web forms, ensuring that test cases accurately reflect real user interactions, and realistic potential scenarios.

Recently, \textit{Large Language Models} (LLMs)~\cite{floridi2020gpt,chang2023survey,zhao2023explainability,zhao2023survey,lemieux2023codamosa} have significantly enhanced \textit{Natural Language Processing} (NLP) technologies~\cite{yang2023harnessing}, leading to a groundbreaking era of \textit{Artificial Intelligence Generated Content} (AIGC)~\cite{chen2024systems}.
Previous studies have shown that LLMs have the potential to improve software engineering~\cite{ozkaya2023application,bano2024large,calo2023leveraging,pearce2023examining,zhang2023automated}.
Due to their reasoning and text-input generation capabilities~\cite{wang2022language,shanahan2024talking,wei2022chain,touvron2023llama,zeng2022glm,yang2023baichuan}, 
LLMs can also use extracted contextual information from complex web-form structures to improve the web-form-test generation process~\cite{schafer2023empirical,yu2023llm}.

Many recent studies have only analyzed a single or very few LLMs~\cite{floridi2020gpt,touvron2023llama,sanderson2023gpt}. 
For instance, Alian et al. \cite{alian2024bridging} evaluated two LLMs (GPT-4~\cite{sanderson2023gpt} and LLaMa2~\cite{touvron2023llama}) in the context of web-form testing. 
They used the Form Entity Relationship Graph (FERG) to model semantic relationships among form elements.
They also incorporated a feedback-driven mechanism to refine input constraints based on real-time form-submission feedback:
This provides valuable insights, and has established an important foundation for applying LLMs in web-form testing.
However, their study was limited to only two LLMs, within a specific framework: Questions about how various LLMs perform across diverse testing scenarios remain. 
For example, it is unclear how LLMs handle different HTML complexities, diverse contextual information, or alternative prompt designs in web-form testing.
To the best of our knowledge, no comprehensive research has systematically evaluated multiple LLMs in the context of web-form testing.
Motivated by these facts, our study analyzes 11 different LLMs, providing a broad perspective on their performance and applicability in automated web-form testing.

\textbf{Our Work:}
(1) We accessed 11 different LLMs using publicly available APIs.
(2) We designed three types of prompts based on the HTML content of web forms for the LLMs: 
\textit{\textbf{R}aw \textbf{H}TML for Task \textbf{P}rompt} (RH-P); \textit{\textbf{L}LM-Processed \textbf{H}TML for Task  \textbf{P}rompt} (LH-P); and \textit{\textbf{P}arser-Processed \textbf{H}TML for Task \textbf{P}rompt} (PH-P).
(3) We generated 14,454 web-form tests and executed them on 30 open-source Java web projects, with 146 web forms, from GitHub to evaluate the capabilities of the LLMs.
We categorized the web forms into five functional types: 
\textit{authentication forms}, \textit{profile forms}, \textit{content management forms}, \textit{search forms}, and \textit{transaction forms}.

\textbf{Key Findings:} 
(1) Different LLMs achieved different \textit{successfully-submitted rates} (SSRs)
---
the proportions of the LLMs-generated web-form tests that could be successfully inserted into the web forms and submitted
---
with GPT-4, GLM-4, and Baichuan2 LLMs performing best.
The model size also appeared to significantly impact the generation performance, such as with the LLaMa2 series of LLMs.
(2) Compared with GPT-4, the other LLMs had more difficulty generating appropriate web-form tests, achieving significantly lower SSRs (between 9.10\% and 74.15\% lower).
Compared with GPT-3.5, some LLMs (such as GLM-3, GLM-4, Baichuan2, and Spark-3.5) were better able to generate appropriate web-form tests, achieving higher SSRs.
(3) Different contextual information and prompt constructions have different impacts on the LLMs' effectiveness, with prompts constructed from parser-processed HTML (PH-P) generally being the best.

\textbf{Practical Implications:} 
(1) Extraction of contextual information from web forms requires fully understanding their properties, and ensuring accurate parsing of the HTML content.
(2) To prune the HTML content of web forms and reduce complexity, redundant components (e.g., the customized components introduced by developers) need to be removed. 
It is also better if it is not necessary to provide complete and complex web-form contextual information.
However, the pruned components should not relate to the key contextual information of the web form (e.g., IDs, which are critical for inserting the generated tests into the corresponding web forms).
(3) When selecting a specific LLM to generate tests for web forms, if there are no practical testing constraints (such as data privacy issues~\cite{wu2024new}), then GPT-4 is preferable (due to its high quality and effective web-form tests).

\textbf{Contributions:} This study offers the following significant contributions:
\begin{itemize}[leftmargin=20pt, labelindent=0pt, itemindent=0pt]
    \item 
    To the best of our knowledge, this is the first empirical study that systematically investigates the potential of multiple (11) LLMs for automated web-form-test generation (going beyond earlier work focused on GPT-4 and LLaMa2).

    \item 
    We prune the HTML to reduce the complexity of the web-form structure. 

    \item
    We propose three context-construction approaches to extract contextual information from web forms for prompt construction.

    \item 
    We select 146 web forms from 30 open-source Java web applications to deeply explore the effectiveness of the different LLMs.
    
    \item 
    We identify three key findings from the experimental results, and provide two practical implications for future research on web-form testing. 
\end{itemize}

The datasets, source code, and experimental results related to this study are available at \url{https://github.com/abelli1024/web-form-test-gen-empirical-study}.

The rest of this paper is organized as follows:
Section~\ref{SEC:Preliminary} introduces some preliminaries, including the HTML parsing process, web-form testing, and LLMs.
Section~\ref{SEC:Study Design} describes the approach taken to conduct the empirical study.
Section~\ref{SEC:Experimental Design} explains the design of our experiments.
Section~\ref{SEC:Experimental Results} presents and analyzes the experimental results, and also discusses some implications and insights for using LLMs in web-form testing.
Section~\ref{SEC:Related Work} outlines some related work.
Section~\ref{SEC:Conclusions and Future Work} concludes the paper and discusses some potential future work.

\section{Preliminaries
\label{SEC:Preliminary}}

This section provides a brief overview of the basic concepts of HTML parsing, web-form testing, and LLMs.

\subsection{HTML Parsing Process}

Listing \ref{lst:html-script} shows an example of a basic HTML structure.
It is made up of various components, including:
title in the \texttt{<head>} (Lines \ref{line:htmlScriptL3} to \ref{line:htmlScriptL5});
content in the \texttt{<body>} (Lines \ref{line:htmlScriptL6} to \ref{line:htmlScriptL9}); and 
image files in the \texttt{<img>} (Lines \ref{line:htmlScriptL8}).
These components provide instructions to the browser for how to display information~\cite{goodman2002dynamic}.

\begin{lstlisting}[caption={The basic HTML structure.}, label=lst:html-script, captionpos=b,escapechar=|]
<!DOCTYPE HTML> 
<html>
  <head>|\label{line:htmlScriptL3}| 
    <title>Sign In.</title> 
  </head>|\label{line:htmlScriptL5}| 
  <body>|\label{line:htmlScriptL6}| 
    <h1 class="ng-personal-signin-title">Sign In.</h1>
    <div class="ng-logo">
        <img src="logo.png" alt="error"/>|\label{line:htmlScriptL8}|
    </div>
    <div class="ng-personal-signin">
        <form action="/signIn" novalidate="" class="gdpr-hidden-personal-signin">
            <label for="username" class="overlabel">Email Address:</label>
            <input id="username" name="username" type="text" class="ng-untouched"/>
            <label for="password" class="overlabel">Password:</label>
            <input id="password" name="password" type="text" class="ng-untouched"/>
            <button type="submit" class="btn-primary">Sign In</button>
        </form>
    </div>
  </body>|\label{line:htmlScriptL9}|
</html>
\end{lstlisting}

\begin{figure*}[ht]
    \centering
    \graphicspath{{Graphs/}}
    \includegraphics[width=0.9\textwidth]{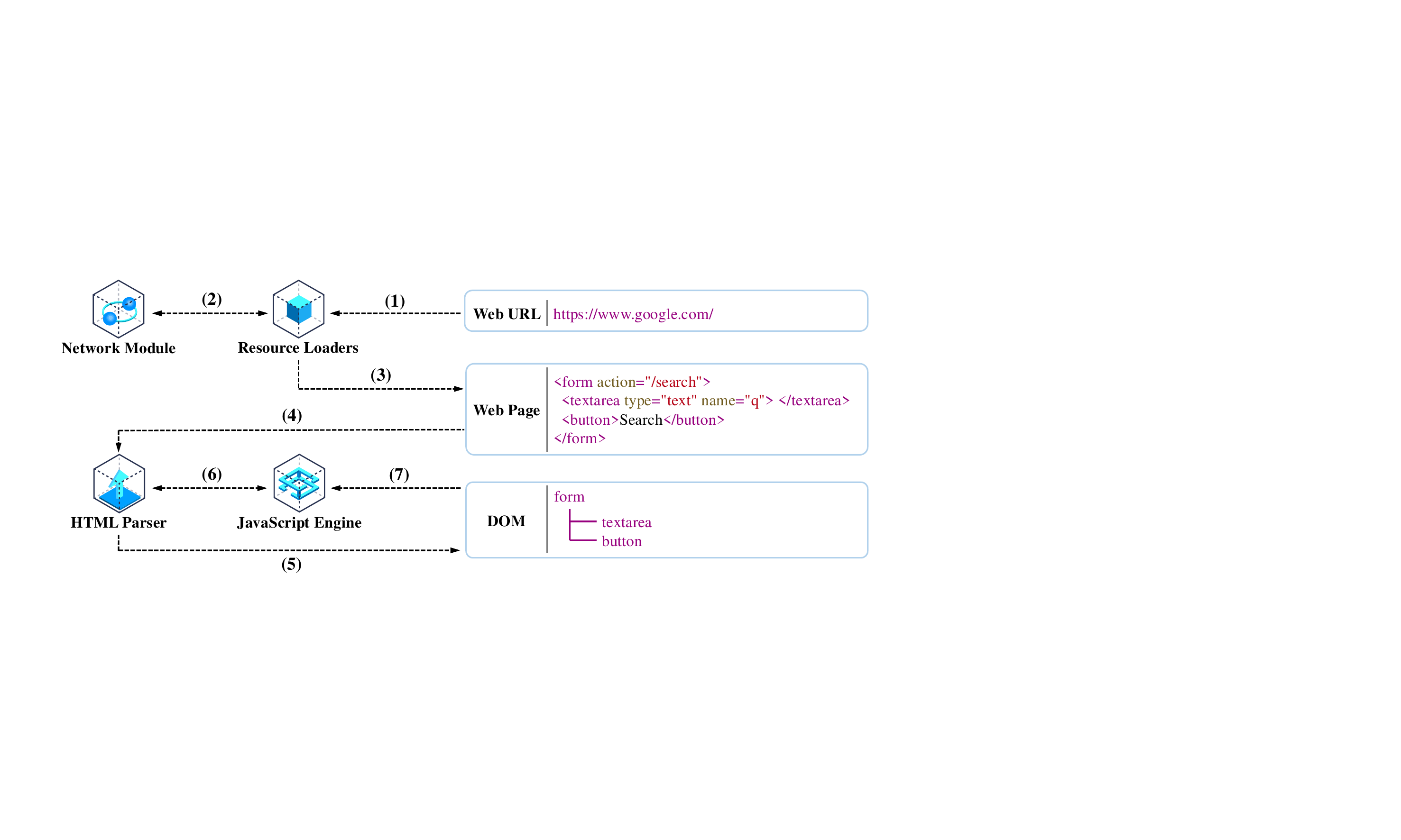}
    \caption{HTML parsing process.}
    \label{FIG: HTML parsing process}
\end{figure*}

The process of parsing an HTML structure is illustrated in Figure \ref{FIG: HTML parsing process}. 
It involves the following steps:
(1) The resource loader is initially activated to load the webpage corresponding to the URL (e.g., https://www.google.com/). 
The loader sends a request to the network module to retrieve the corresponding web content.
(2) The loader then uses the network module to initiate requests and handle responses:
It retrieves the raw HTML content of the requested page, which is then forwarded for further processing.
(3) Data can be obtained from web pages or resources through synchronous and asynchronous methods:
Static content is directly available in the initial response, but dynamic content requires additional network requests to retrieve external dependencies or API-driven.
(4) The web page is then passed to an HTML interpreter and transformed into a series of words or tokens that represent the structural components of the document, and serve as the basis for subsequent parsing.
(5) Based on these tokens, the interpreter constructs nodes, forming a DOM tree~\cite{goodman2002dynamic}.
These nodes define the structure and relationships of the document elements, serving as the foundation for rendering and script execution.
For example, a <form> element in the HTML source and its child nodes (e.g., <textarea> and <button>) are parsed into corresponding DOM nodes.
(6) If a node is written in JavaScript, then the JavaScript engine is called to interpret and execute it, allowing dynamic modifications to the DOM tree, and interaction with other elements of the webpage.
(7) JavaScript code may modify the structure of the DOM tree by dynamically adding, removing, or updating elements.  
Additionally, if a node requires external resources (such as images, \textit{Cascading Style Sheets} (CSS), videos, etc.), then the resource loader retrieves them.

\subsection{Web-Form testing
\label{SEC:Web form Testing}}

This section presents some background information about the basic web-form testing framework, and some strategies for identifying the web elements' locations.

\subsubsection{Basic Web-Form Testing Framework}
Web-form testing ensures the interactive functionality, usability, and compatibility of web forms in a web application.
Automated testing methods have become widely used~\cite{alonso2022arte,leotta2016approaches,garcia2021automated}:
Selenium~\cite{selenium2024}, for example, is a popular automated testing framework for web-form testing.
The Selenium WebDriver, a core part of Selenium~\cite{garcia2021automated,seleniumWebDriver2024}, allows testers to automatically control web applications' behavior on real browsers, using automated test scripts~\cite{gundecha2018selenium}.
It uses the native automation support from web browsers~\cite{vila2017automation} to enable an end-to-end test execution~\cite{leotta2016approaches,leotta2023challenges}.

\begin{figure*}[!t]
    \centering
    \graphicspath{{Graphs/}}
    \includegraphics[width=0.98\textwidth]{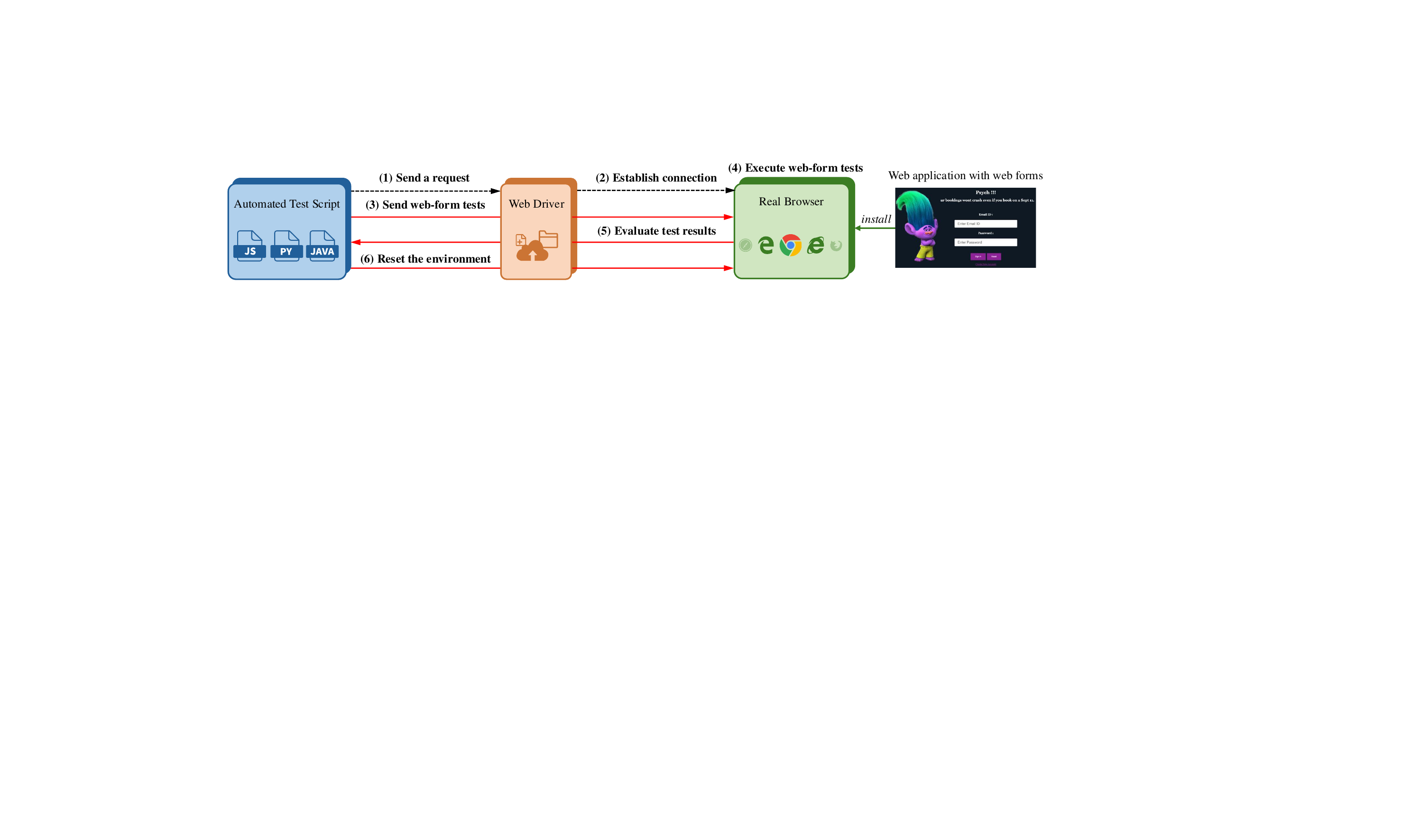}
    \caption{Process of using Selenium to control automated testing (black dashed lines represent standard steps; red solid lines are steps requiring WebDriver).}. 
    \label{FIG:Selenium WebDriver controlling Chrome}
\end{figure*}

Figure \ref{FIG:Selenium WebDriver controlling Chrome} outlines the main steps in web-form testing using Selenium. 
The process includes both standard steps (shown with black dashed lines) and WebDriver-dependent steps (highlighted with red solid lines).
In Step (1), a customized test script is developed that includes the execution logic of the testing task, and the initiation request for the Selenium WebDriver.
In Step (2), WebDriver sends an {\em establish connection} command to the browser where the web application with the \textit{Application Under Test} (\textit{AUT}) has been installed.
At this point, the pipeline between the test script and the web application has been established.
In Step (3), using WebDriver, the UI operations for the web form (e.g., clicking a button or filling out a text field) are defined in the test script and sent to the browser.
In Step (4), the web-form tests are executed within the AUT (the web form) on the browser, accompanied by a series of relevant actions (such as clicking buttons, triggering submission, etc.).
After the tests, in Step (5), the browser returns the results to the testers through WebDriver.
Finally, in Step (6), using WebDriver, the testing environment is reset to its original state.

Web-form testing can be complicated due to two properties of web forms: 
(1) Web forms can comprise various elements, including both basic and customized components. 
For example:
\textit {tags} are the basic components that specify the input content type;
\textit {elements} establish the web form's layout; and 
\textit {attributes} specify the functionality and characteristics of the elements.
In addition, the customized components introduced by developers can enhance the diversity of the web-form structures and include some control logic.
(2) Web forms also provide diverse contextual information for users to interact with the web applications:
For example, web forms may provide different approaches for users to select from alternative options, such as using a selector list or a group of radio buttons.

While frameworks such as Selenium WebDriver are commonly used, they have limitations when handling complex web-form components. 
Many web forms include customized input types and require dynamic validation, making effective testing difficult. 
LLMs have the potential to address these challenges~\cite{wang2024software}: 
LLMs can understand and simulate complex user interactions, and can handle dynamic and customized form elements, enabling the generation of effective web-form tests.

\subsubsection{Web-Element Location Identification Strategies}
In web-form testing, it is necessary to identify and locate the web elements before interacting with them.
Selenium WebDriver provides eight strategies that use attributes of the web elements to locate them (\textit{tag name}, \textit{element ID}, \textit{CSS selector}, \textit{element name}, \textit{class name}, \textit{link text}, \textit{partial link text}, and \textit{XPath}) \cite{seleniumLocatorStrategies2024}.
This study uses the following three strategies:
\begin{itemize}
    \item[(1)] The \textit{tag-name-based strategy} identifies the location of elements whose tag name (including \texttt{div} and \texttt{input}) matches the search value.

    \item[(2)] he \textit{element-ID-based strategy} gets the location of elements whose ID attribute matches the search value.

    \item[(3)] The \textit{CSS-selector-based strategy} locates elements matching a CSS selector defined in the component.
\end{itemize}

\subsection{LLMs\label{SEC:LLM}}

LLMs are complex deep neural networks trained on various datasets, including books, code, articles, and websites. 
This training enables the model to discern and replicate the intricate patterns and relationships inherent in the language it learns. 
As a result, LLMs can produce coherent content ranging from grammatically accurate text to syntactically correct code snippets~\cite{ozkaya2023application}.
LLMs~\cite{radford2019language,vaswani2017attention} are advanced technologies in deep learning and NLP, with models including GPT~\cite{sanderson2023gpt}, GLM~\cite{zeng2022glm}, and LLaMa2~\cite{touvron2023llama}. 
These models learn complex features (such as language structure, grammar, and semantics) by training on large-scale text data. 
With this training, LLMs can complete more complex and diverse NLP tasks, such as text generation~\cite{li2022pretrained,liu2023fill}, translation~\cite{huang2023towards}, and AI assistance (e.g., contributing to a range of software engineering tasks, including specification generation and the translation of legacy code)~\cite{ozkaya2023application}.

LLMs are used in four key stages of the software engineering lifecycle: 
1) Software requirements and design (e.g., software specifications generation~\cite{xie2023impact} and GUI layouts~\cite{brie2023evaluating});
2) Software development (e.g., code generation~\cite{li2023acecoder,zhang2023self} and code summary~\cite{rukmono2023achieving,sun2020automatic});
3) Software testing (e.g., unit test generation~\cite{tufano2022generating,schafer2023empirical,Chen2024} and GUI testing~\cite{liu2023fill,liu2024make}); and 
4) Software maintenance (e.g., code review~\cite{li2022automating,li2022auger} and bug-report detection~\cite{zhang2023cupid,plein2023can}). 
LLMs can solve complex software engineering problems and promote better software engineering development.

\begin{figure*}[!b]
    \centering
    \graphicspath{{Graphs/}}
    \includegraphics[width=\textwidth]{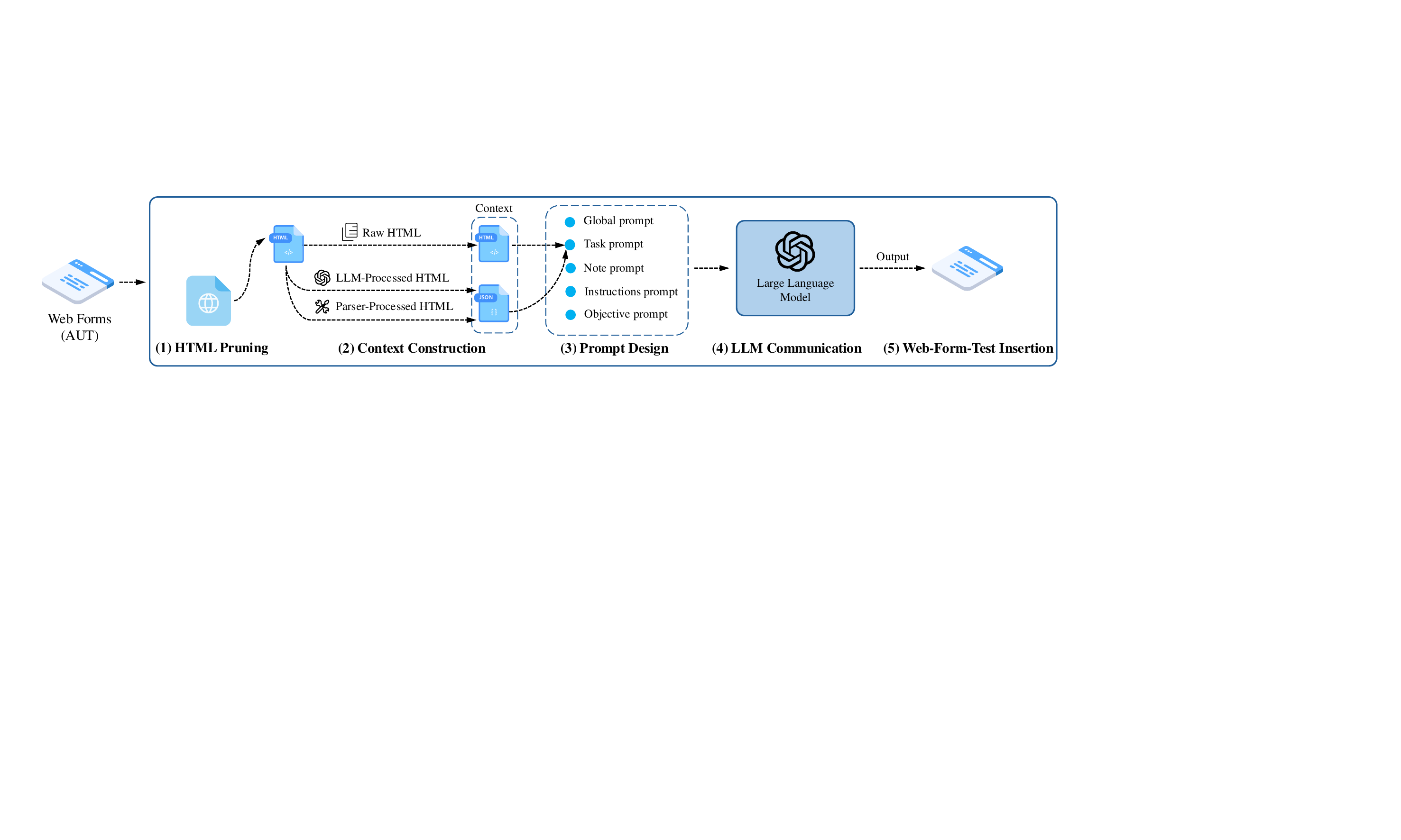}
    \caption{Framework of this empirical study.}
    \label{FIG: framework}
\end{figure*}

\section{Approach
\label{SEC:Study Design}}

This study focuses on automated web-form-test generation for web forms. 
Figure \ref{FIG: framework} illustrates the framework of this empirical study, which includes the following five steps: 
1) \textit{HTML pruning};
2) \textit{context construction};
3) \textit{prompt design};
4) \textit{LLM communication}; and
5) \textit{web-form-test insertion}.

\subsection{HTML Pruning}
This section outlines the method of HTML pruning, a critical technique for clarifying the contextual information of web forms. 

Although HTML has a significant tree-like DOM structure~\cite{goodman2002dynamic}, it can also contain various types of components, which can make it difficult to filter out the key ones. 
Web-form developers generally use essential basic components, such as \textit{name} and 
\textit{ID}\footnote{For ease of description, unless explicitly stated otherwise, in this paper the \textit{name} represents the \textit{element name} and \textit{ID} represents the \textit{element ID}}. 
Some customized components can also be added to make the code logic more transparent, such as \textit{data-ID} and \textit{user-role}.
However, these customized components may make the HTML pruning process more complex.
Therefore, designing a method for pruning HTML that systematically retains critical components while removing non-essential ones is becoming increasingly challenging.

After analysis of the web-form structure, some key components (\textit{ID}, \textit{name}, \textit{type}, \textit{placeholder}, \textit{src}, \textit{disabled}, \textit{readonly}, \textit{value}, \textit{checked}, \textit{required}, \textit{selected}) are retained while the remaining components get pruned.
In other words, the HTML contents of the web forms are simplified while maintaining semantic integrity.
This makes the entire HTML clear, and the web form readable.
Because some customized components (such as \textit{data-ID} and \textit{user-role}) do not change the logic or semantic integrity of the web forms, they are removed during the pruning process.
However, the key components (which are fundamental to the functionality and operability of web forms) must be retained:
For example, \textit{ID} and \textit{name} are needed to identify the web elements within the web forms, and are inserted with the LLM-generated tests.
(More details will be discussed in Section \ref{sec: Web-Form-Test Insertion}.)

\begin{algorithm}[!t]
    \DontPrintSemicolon
    \footnotesize
    \caption{HTML Pruning $\texttt{\textbf{pruneHTML(}}\mathcal{D}, \mathcal{S}\texttt{\textbf{)}}$}
    \label{ALG:Html pruning}
      \SetKwData{Left}{left}\SetKwData{This}{this}\SetKwData{Up}{up}
      \SetKwInOut{Input}{Input}\SetKwInOut{Output}{Output}
      \renewcommand{\algorithmicendwhile}{\algorithmicend~\algorithmicwhile}

    \Input{WebDriver $\mathcal{D}$, Filter tag set $\mathcal{S}$.}
    \Output{Pruned HTML $\mathcal{H}$.}
    $\mathcal{F} \leftarrow \texttt{\textbf{findElements(}}\mathcal{D},  form\texttt{\textbf{)}}$;

    $\mathcal{H} \leftarrow \{\}$;
    
    \For{each~element $\chi\in \mathcal{F}$}
    {
        \For{each~attribute $\alpha \in \chi$}
        {
            $\tau \leftarrow \texttt{\textbf{getTag(}}\alpha\texttt{\textbf{)}}$;
            
            \If{$\tau \notin \mathcal{S}$}
            {
                $\chi \leftarrow \texttt{\textbf{remove(}}\alpha,\chi\texttt{\textbf{)}}$;
            }
        }
        $\mathcal{H} \leftarrow \textbf{\texttt{append(}}\mathcal{H}, \chi \textbf{\texttt{)}}$;
    }
    \textbf{return} $\mathcal{H}$;
\end{algorithm}

Algorithm \ref{ALG:Html pruning} provides the pseudocode for HTML pruning, where $\mathcal{D}$ is a WebDriver, and $\mathcal{S}$ is a filter tag set.
In the initialization stage, a web-form $\mathcal{F}$ is generated by searching for elements from $\mathcal{D}$ with the keyword $form$ ($\texttt{\textbf{findElements(}}\mathcal{D}, form\texttt{\textbf{)}}$).
For each element $\chi$ in $\mathcal{F}$, the algorithm checks each attribute $\alpha$ to find the potential tag $\tau$ ($\texttt{\textbf{getTag(}$\alpha$\textbf{)}}$).
If the tag $\tau$ does not belong to the tag set $\mathcal{S}$, then the attribute $\alpha$ is removed from the element $\chi$; 
otherwise, $\alpha$ is retained. 
Finally, the updated element $\chi$ is appended to a pruned HTML $\mathcal{H}$.

We use a specific example to illustrate the pruning process:
Listing~\ref{lst:before-sample-html-pruning} presents the original HTML structure of a web form before applying the pruning process.
In this structure, attributes such as \textit{id}, \textit{name}, and \textit{type} are essential for identifying and interacting with the web elements, making them critical components that must be retained. 
In contrast, attributes like \textit{data-ID} and \textit{user-role} do not contribute to the semantic integrity or operational functionality, and are therefore removed during pruning.
After applying the pruning algorithm with the filter tag set $\mathcal{S} = \{\textit{id}, \textit{name}, \textit{type}, \textit{placeholder}, \textit{src}, \textit{disabled}, \textit{readonly}, \textit{value}, \textit{checked}, \textit{required}, \textit{selected}\}$,
Listing~\ref{lst:after-sample-html-pruning} presents the simplified HTML structure of the web form:
\begin{lstlisting}[caption={The HTML structure of the web form before pruning.}, label=lst:before-sample-html-pruning, captionpos=b,escapechar=|]
<form>
  <input id="username" name="user_name" type="text" data-id="1234" class="text form-control"/>
  <input id="password" name="user_pass" type="password" user-role="admin" class="text form-control"/>
</form>
\end{lstlisting}
\begin{lstlisting}[caption={The HTML structure of the web form after pruning.}, label=lst:after-sample-html-pruning, captionpos=b,escapechar=|]
<form>
  <input id="username" name="user_name" type="text"/>
  <input id="password" name="user_pass" type="password"/>
</form>
\end{lstlisting}
This example shows how the pruning process removes non-essential attributes while retaining critical ones, simplifying the HTML structure.

\subsection{Context Construction
\label{SEC: Context Construction}}

This section offers an overview of the context-construction method, which involves transforming the web form's contextual information into task prompts that can guide the LLM's generation of web-form tests.
Here, the contextual information refers to the structural elements and attributes
---
derived from web-form HTML
---
as well as their interrelations.
LLMs have demonstrated a strong logical understanding~\cite{cochran2023improving,kumar2024large}, and an ability to generate high-quality content~\cite{sanderson2023gpt}. 
However, this is strongly dependent on the prompt quality~\cite{wei2022chain}.
Building the LLM prompts requires that the contextual information be constructed from the web form's HTML content, which can include various components, such as \textit{placeholders} and \textit{attributes}.

We propose three methods of context construction based on the web-form structure: 
\textit{Raw HTML}, \textit{LLM-Processed HTML}, and \textit{Parser-Processed HTML}. 
As shown in Figure~\ref{FIG: framework}, context construction is a core element of our framework.
Each of the three proposed methods represents a distinct approach to extracting and constructing the contextual information: 
The detailed step-by-step processing of each approach is further illustrated in the following sections.

\subsubsection{Context from Raw HTML}
The first method constructs context directly from the raw HTML
--- 
this was motivated by the fact that the HTML layout effectively represents the relationships among the various components.
An advantage of this method is its simplicity:
It directly uses the raw HTML without any additional processing or conversion.
Due to the limited amount of context that LLMs can handle~\cite{arefeen2024leancontext}, it is necessary to \textit{simplify} the raw HTML content by removing some CSS and customized components defined by developers 
---
this is done by the $\texttt{\textbf{pruneHTML(}}\mathcal{D}, \mathcal{S}\texttt{\textbf{)}}$ function.
The pruned HTML is then reorganized as a string, which is considered to be the contextual information.

To construct the prompt from the pruned HTML, the retained elements—including form fields, labels, and relevant attributes—are directly embedded into a predefined prompt structure. 
This ensures that the LLM receives only essential contextual information while maintaining the structural relationships necessary for understanding the form's functionality.

\subsubsection{Context from LLM-Processed HTML} 
The second method of context construction uses the LLM to pre-process the HTML. 
An advantage of this method is its ability to handle more complex web-form structures, utilizing the LLM’s understanding to capture the intricate connections among form elements.
The LLM parses the contextual information from the HTML content (which includes accurately assembled elements of the web form, such as input tag name, tag ID, and tag type).
In general, LLMs convert web-form HTML into a list of \textit{JavaScript Object Notation} (JSON) structured results.
The context is then constructed by traversing the web-form JSON structure, extracting the corresponding JSON information, and concatenating it into the context content.

Algorithm \ref{ALG:LLM-Processed HTML} provides the pseudocode for extracting context from the LLM-Processed HTML. 
Initially, $s$ is an empty string, and the pruned HTML $\mathcal{H}$ is generated by $\texttt{\textbf{pruneHTML(}}\mathcal{D}, form\texttt{\textbf{)}}$. 
$\mathcal{H}$ is then transformed into a JSON-structured objects list $\Omega$ using a specific LLM ($\texttt{\textbf{parseLLM(}}\mathcal{H}, \mathcal{L}\texttt{\textbf{)}}$). 
For each element $\chi \in \Omega$, the algorithm uses the following functions to capture the contextual information:
$\texttt{\textbf{getName(}}\chi\texttt{\textbf{)}}$ retrieves the name of the JSON object $\chi$ (e.g., ``hint text''); 
$\texttt{\textbf{getValue(}}\chi\texttt{\textbf{)}}$ obtains the corresponding value of $\chi$ (e.g., ``Please enter your name.''); and
$\texttt{\textbf{getContext}}$ concatenates the obtained name and value as natural language sentences (e.g., ``The hint text is `Please enter your name.{'}'').
Then, all the reconstructed HTML information is reformed into a parsed context string as the context.
This method processes web-form HTML into a structured JSON format, preserving essential attributes (e.g., input tag names, tag IDs, and tag types) while removing irrelevant or unnecessary elements.  
The resulting structured contextual information enables the LLM to interpret web-form components more precisely, ensuring accurate test generation by maintaining explicit relationships among input fields and their attributes.

\begin{algorithm}[!t]
    \DontPrintSemicolon
    \footnotesize
    \caption{Context from LLM-Processed HTML}
    \label{ALG:LLM-Processed HTML}
      \SetKwData{Left}{left}\SetKwData{This}{this}\SetKwData{Up}{up}
      \SetKwInOut{Input}{Input}\SetKwInOut{Output}{Output}
      \renewcommand{\algorithmicendwhile}{\algorithmicend~\algorithmicwhile}

    \Input{WebDriver $\mathcal{D}$, Filter tag set $\mathcal{S}$, LLM $\mathcal{L}$.}
    \Output{Context string $s$.}
    
    $s \leftarrow \varepsilon$;~~~~~~~~~~~~~~~~~// $s$ is initialized as an empty string.

    $\mathcal{H} \leftarrow \texttt{\textbf{pruneHTML(}}\mathcal{D}, \mathcal{S}\texttt{\textbf{)}}$;

    $\Omega \leftarrow \texttt{\textbf{parseHTML(}}\mathcal{H}, \mathcal{L}\texttt{\textbf{)}}$;~~~~~// Parse the HTML as a list of JSON structured results by the LLM.
    
    \For{each $\chi \in \Omega$} 
    {\label{line:LLMProcessed4}
        $\tau \leftarrow \texttt{\textbf{getContext(}}\texttt{\textbf{getName(}}\chi\texttt{\textbf{)}},\texttt{\textbf{getValue(}}\chi\texttt{\textbf{)}}\texttt{\textbf{)}}$;
        
        $s \leftarrow \texttt{\textbf{append(}}s,\tau\texttt{\textbf{)}}$;
    }\label{line:alg2:forend}
    \textbf{return} $s$;
\end{algorithm}

\subsubsection{Context from Parser-Processed HTML}
When evaluating the effectiveness of the LLM-processed HTML approach, we found that LLMs occasionally fail to process HTML into JSON objects.
This is because of the non-determinism of LLMs~\cite{Ouyang23}, resulting in the omission of key contextual information for the web-form tests.
This issue will be discussed in detail in Section~\ref{sec: results-rq1}.
To avoid missing any key contextual information, we propose a third context-construction method.
This method is similar to the second, with the only difference being the use of different HTML-parsing functions ($\texttt{\textbf{parseHTML}}$ in Algorithm \ref{ALG:LLM-Processed HTML}):
Instead of an LLM as the HTML parser, a Java-based HTML parser (Jsoup\footnote{\url{https://jsoup.org/}.}) converts the web-form HTML into a list of JSON-structured results.
The context is also constructed as a string.
This method provides a structured and deterministic extraction of web-form attributes, mitigating the non-deterministic outputs that may occur in LLM-based parsing.
The advantage of this method is its ability to reliably parse the complex relationships within web forms, ensuring complete and accurate extraction of relevant information.

\subsection{Prompt Design
\label{SEC:Prompt Design}}

This section provides a detailed introduction to the process of prompt design. 
The constructed prompts are directly used to guide the LLM in the web-form-test generation.

After collecting and constructing the context (as discussed in Section \ref{SEC: Context Construction}), the next stage is to design the prompts for the LLMs. 
This includes five steps that produce the following five prompt types:
1) \textit{global prompt};
2) \textit{task prompt}; 
3) \textit{note prompt}; 
4) \textit{instruction prompt}; and
5) \textit{objective prompt}.
These prompt types were inspired by previous studies~\cite{liu2024make, cui2024large} and have been adjusted for web-form-test generation tasks.
Once the five prompt types have been constructed, they are concatenated into a string as the final prompt for the LLMs.
Figure \ref{FIG: prompt-structure} presents the basic framework of the prompt structure, where the \textit{``\%s''}s are replaced by the extracted information when testing the AUTs.

\subsubsection{Global Prompt}
The global prompt contains the most basic information, such as the name of the web application, the title of the web form, and the number of elements within the form.
It establishes the basic context for the task.

\subsubsection{Task Prompt}
\begin{figure*}[!t]
    \centering
    \graphicspath{{Graphs/}}
    \includegraphics[width=\textwidth]{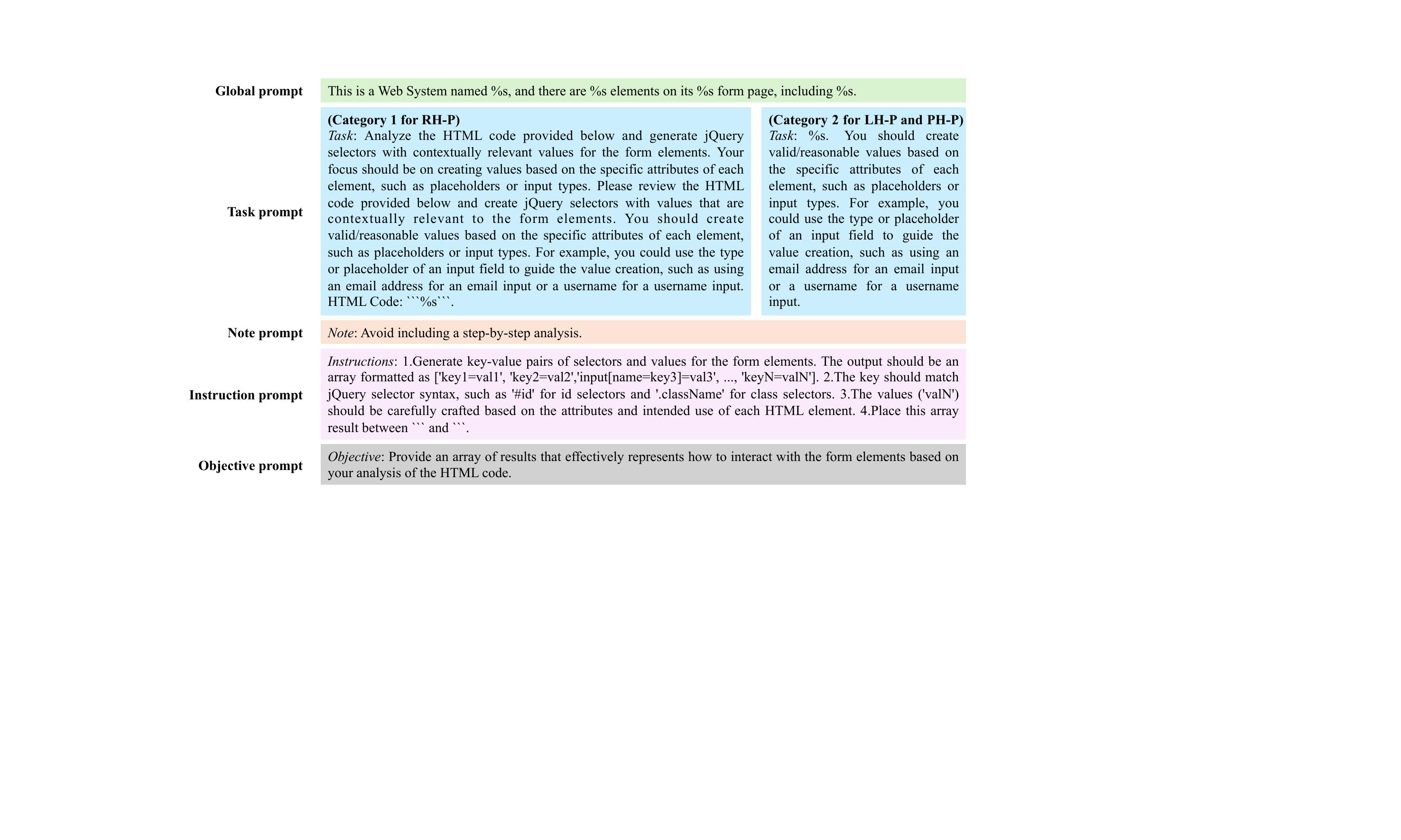}
    \caption{Basic framework of the prompt structure.}
    \label{FIG: prompt-structure}
\end{figure*}

The task prompt includes the main contextual information for the web form.
Based on the three types of context (Section \ref{SEC: Context Construction}), three types of task prompts are created:
\textit{\textbf{R}aw \textbf{H}TML for Task \textbf{P}rompt} (RH-P); 
\textit{\textbf{L}LM-Processed \textbf{H}TML for Task \textbf{P}rompt} (LH-P); and 
\textit{\textbf{P}arser-Processed \textbf{H}TML for Task \textbf{P}rompt} (PH-P). 
Two categories of natural language sentences are designed for the LLMs to analyze: 
Category 1 asks the LLM to analyze the structure of the HTML code; and 
Category 2 asks the LLM to analyze the contextual information in natural language expressions.
As shown in Figure \ref{FIG: prompt-structure}, 
RH-P uses Category 1, while LH-P and PH-P use Category 2 (as they are based on the contextual information in the natural language expressions).

\subsubsection{Note Prompt}
The note prompt guides the LLM toward generating more concise outputs by restricting the inclusion of unnecessary details. 
It prioritizes brevity over elaboration, ensuring that responses focus on delivering essential information.
For instance, while step-by-step explanations may be appropriate in other contexts, they can introduce excessive detail that may hinder the efficient extraction of relevant content.

This could make it more difficult to extract key information efficiently, potentially leading to less clear and more verbose outputs. 
By emphasizing brevity and precision, the note prompt helps streamline responses, reducing unnecessary variations and improving overall clarity.

\subsubsection{Instruction Prompt}
The instruction prompt enforces a consistent and structured format in the outputs generated by the LLM, ensuring that they align with predefined requirements.
The outputs must be formatted as an array of key-value pairs (\texttt{[``key1=val1'', ``key2=val2'', ``input[name=key3]=val3'', ..., ``keyN=valN'']}), where each key represents a jQuery selector (e.g., \#id for ID selectors and .className for class selectors), and each value corresponds to attributes of the associated HTML element.
To preserve the expected structure, the entire array must be enclosed in a pair of triple quotation marks to ensure clarity and consistency.

\subsubsection{Objective Prompt}
The objective prompt instructs the LLM to strictly adhere to the defined requirements and generate appropriate results for the form elements. 
The generated results must be formatted as an array that correctly represents the relationships among selectors and their corresponding values.
This requires identifying each form element’s attributes and intended functionalities, thereby ensuring that the output is syntactically precise. 
By enforcing strict alignment with the structural and functional aspects of HTML, the objective prompt guarantees that the outputs are reliable, actionable, and directly applicable in automated web-form testing. 
Furthermore, this focus on precision reduces ambiguities and promotes consistency, ultimately facilitating the accurate validation of form interactions.

\begin{figure*}[!t]
    \centering
    \graphicspath{{Graphs/}}
    \includegraphics[width=0.99\textwidth]{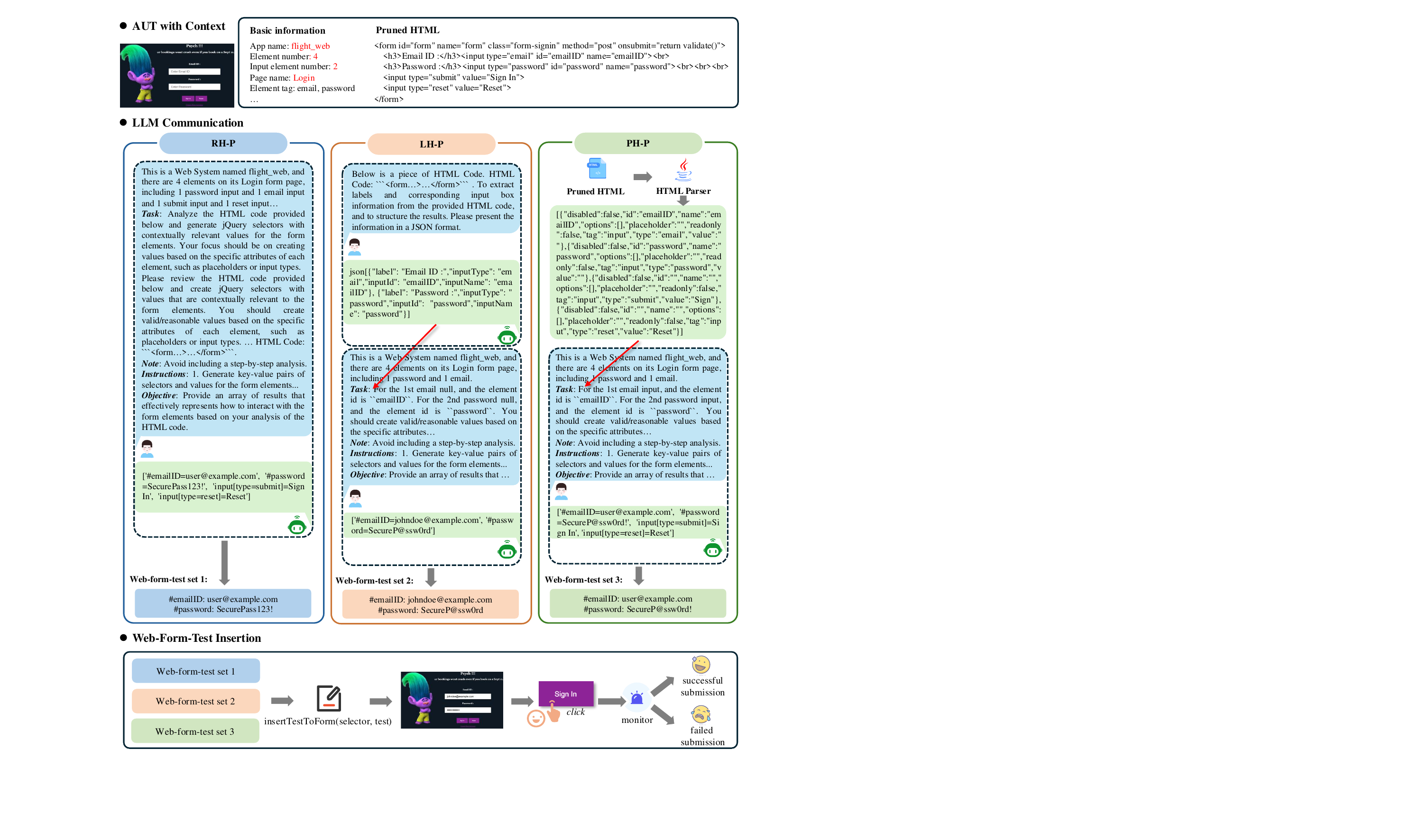}
    \caption{Example illustrating the entire web-form testing process.}
    \label{FIG: example-llms}
\end{figure*}

\subsection{LLM Communication
\label{SEC:LLM Communication}}

This section elaborates on the communication with the LLM, which is a critical step for the LLMs to generate web-form tests.
Once the communication connection with the LLM is established, the LLM can be used to generate the web-form tests based on the extracted contextual information.

First, we standardize the encapsulation of various LLM APIs, which leads to better management, easier communication, and more convenient collection of experimental data.
We can communicate with a specific LLM by sending the name of the  LLM to this encapsulated API.
Second, when running the AUT, we use three types of task prompts to construct the complete prompt (Section \ref{SEC:Prompt Design}).
This complete prompt then guides the LLM to generate the web-form tests:
The test script automatically sends the prompts as a message.
Figure \ref{FIG: example-llms} shows the contextual information being concatenated into three types of prompts (RH-P, LH-P, and PH-P), and then passed to a specific LLM to generate the web-form tests.
The web-form tests are then extracted from the response text, and converted into a set of key-value pairs for web-form-test insertion: 
The key stores the \textit{selector} of each component (i.e., its location \cite{seleniumLocatorStrategies2024}); 
and the value stores the corresponding generated web-form test.

To enhance the integration of LLM-generated web-form tests into the automated testing process, we implemented a structured storage and execution pipeline. 
Once the key-value pairs representing the web-form test inputs are extracted, they are systematically stored in a relational database (e.g., MySQL) to facilitate efficient test execution, reproducibility, and comparative analysis. 
The database schema maintains key metadata --- including web URL, web-form content, the LLM model used, input prompts, extracted test values, and execution outcomes --- allowing for traceability and systematic evaluation of LLM performance across different web forms.

\subsection{Web-Form-Test Insertion
\label{sec: Web-Form-Test Insertion}}

This section explains the insertion process of the web-form tests generated by the LLM. 
This step is the key to achieving automation of the entire web-form testing process.

After extracting all the LLM-generated web-form tests, they are stored in a set with the corresponding selectors for all the web-form components.
If the selector detects the web-form component, the corresponding generated web-form test is inserted.
As shown in Figure \ref{FIG: example-llms}, the function \texttt{\textbf{insertTestToForm(}}\textit{selector}, \textit{test}\texttt{\textbf{)}} implements the component-selection and web-form-test insertion process.
Listing \ref{lst:insertTestToForm} presents an example Java implementation of the \texttt{\textbf{insertTestToForm(}}\textit{selector}, \textit{test}\texttt{\textbf{)}} function.
The functions invoked in this implementation are based on the Selenium tool~\cite{seleniumWebDriver2024}.
Specifically, as mentioned in Section \ref{SEC:LLM Communication}, the LLMs return web-form tests containing (\textit{selectors}) and the corresponding input strings (\textit{tests}).
First, the \texttt{\textbf{findElement}} function retrieves specific web-form components according to the selector (Line \ref{line:insertTestToForm3}).
Then, the \texttt{\textbf{sendKeys}} function inserts the components with the test string (Line \ref{line:insertTestToForm5}).
No part of the insert process requires any human intervention.

Once the web-form tests are inserted into the corresponding components, some UI operations (e.g., clicking a button) are performed to submit the inserted values to the web application's server.
As illustrated in Figure \ref{FIG: example-llms}, the ``Sign in'' button is automatically clicked to submit the inserted values.
Finally, a monitor is set to wait until the submission triggers a response from the application server:
If a response is caught, then it is considered a successful submission; otherwise, it has failed.

\begin{lstlisting}[language=Java,caption={A Java implementation example of the \texttt{\textbf{insertTestToForm(}}\textit{selector}, \textit{test}\texttt{\textbf{)}} function.}, label=lst:insertTestToForm, captionpos=b,escapechar=|]
public void insertTestToForm(String selector, String test) {|\label{line:insertTestToForm1}| 
    // Find WebElement based on the selector
    WebElement webElement = driver.findElement(By.cssSelector(selector));|\label{line:insertTestToForm3}| 
    // Use the sendKeys method to input the test value
    webElement.sendKeys(test);|\label{line:insertTestToForm5}| 
}
\end{lstlisting}

\section{Study Design
\label{SEC:Experimental Design}}

This section introduces the details of our study design, including the research questions, the selection of LLMs, the evaluation setup, and the experiment environment.

\subsection{Research Questions}
Our research aimed to evaluate the effectiveness of LLMs in generating tests for web forms, guided by the following two research questions:
\begin{description}[leftmargin=0pt, labelindent=0pt]
    \item 
    \textbf{RQ1}: How effective are the web-form tests generated by different LLMs?
    \begin{itemize}[leftmargin=20pt, labelindent=0pt, itemindent=0pt]
        \item 
        \textbf{RQ1.1}: How well do different types of prompts guide the web-form-test generation?
        
        \item 
        \textbf{RQ1.2}: How well do different LLMs perform in generating web-form tests?
    \end{itemize}
    For \textbf{RQ1}, we conducted a series of experiments, with 11 different LLMs, to investigate the effectiveness of different LLM-generated web-form tests from the perspectives of the three prompt types. 
    
    \item 
    \textbf{RQ2}: What is the quality of the generated web-form tests?
    \begin{itemize}[leftmargin=20pt, labelindent=0pt, itemindent=0pt]
        \item 
        \textbf{RQ2.1}: Why are some generated web-form tests not successfully submitted?
        
        \item 
        \textbf{RQ2.2}: What is the quality of the generated web-form tests, from the perspective of testers?
    \end{itemize}
    For \textbf{RQ2}, we analyzed the reasons why some generated web-form tests could not be submitted successfully.
    We also used an online user study to evaluate the quality of the different LLM-generated web-form tests, from the users' perspective.
\end{description}

\subsection{LLM Selection}
After an in-depth investigation and analysis of the current mainstream and widely-used LLMs~\cite{naveed2023comprehensive,chang2023survey,zhao2023survey,min2023recent,wang2024software,kumar2024large}, we selected 11.
These state-of-the-art LLMs were chosen based on the number of parameters~\cite{zhang2024scaling}, the ease of API access and integration~\cite{sparkapi2024}, and the availability of commercial and open-source options~\cite{touvron2023llama,zeng2022glm,yang2023baichuan}. 
Table \ref{TAB:llm-selection} lists some relevant information for these 11 LLMs, including the model name and version, the owning company or organization, the number of parameters, the source website, and the year of its release.
Interestingly, some information about these LLMs was not fully disclosed, such as the number of parameters.
These models were selected to ensure diversity in functionality and applicability, while also considering flexibility and accessibility.
For instance, proprietary models like GPT-4 excel in complex reasoning and text-generation tasks~\cite{sanderson2023gpt}, while open-source alternatives such as LLaMa2 enable customization for diverse research scenarios~\cite{touvron2023llama}.
When these models were released, their capabilities were often vigorously promoted, which makes the empirical research in this article more important and meaningful.

For the parameter settings of the selected LLMs, we adopted the default hyperparameters provided by their respective developers to ensure consistency and reproducibility. 
For instance, OpenAI's GPT models were configured with default values such as temperature (defaults to 1) and top\_p (defaults to 1)\footnote{OpenAI Platform \url{https://platform.openai.com/docs/}.}. 
These settings were extensively tested, and were recommended to balance creativity and coherence in text generation. 
Using default configurations also provides a reliable baseline for evaluating model performance without the variability introduced by manual parameter adjustments.

\begin{table*}[!t]
\caption{List of subject LLMs.}
\label{TAB:llm-selection}
\scriptsize
\centering
\setlength\tabcolsep{1mm} 
\begin{tabular}{@{}c|c|c|c|c|c|c@{}}
\hline
\textbf{No.} & \textbf{Name}   & \textbf{Version}       & \textbf{Company/Organization}        & \textbf{No. of Parameters} & \textbf{Source}                         & \textbf{Year of Release} \\ \hline
1   & GPT-3.5     & gpt-3.5-turbo & OpenAI              & 175B        & \url{https://platform.openai.com}    & 2022 \\
2   & GPT-4       & gpt-4         & OpenAI              & Not Reported          & \url{https://platform.openai.com}    & 2023 \\
3   & GLM-3       & GLM-3         & Zhipuai & Not Reported          & \url{https://www.zhipuai.cn}         & 2023 \\
4   & GLM-4       & GLM-4         & Zhipuai & Not Reported          & \url{https://www.zhipuai.cn}         & 2024 \\
5   & GLM-4V      & GLM-4V        & Zhipuai & Not Reported         & \url{https://www.zhipuai.cn}         & 2024 \\
6   & Baichuan2   & Baichuan2-53B & Baichuan-inc        & 53B       & \url{https://api.baichuan-ai.com}   & 2023 \\
7   & LLaMa2(7B)  & LLaMa2-7b     & Meta                & 7B        & \url{https://replicate.com/} & 2023 \\
8   & LLaMa2(13B) & LLaMa2-13b    & Meta                & 13B       & \url{https://replicate.com/} & 2023 \\
9   & LLaMa2(70B) & LLaMa2-70b    & Meta                & 70B       & \url{https://replicate.com/} & 2023 \\
10  & Spark-3     & Spark-3       & IFLYTEK             & Not Reported          & \url{https://spark-api.xf-yun.com}   & 2023 \\
11  & Spark-3.5   & Spark-3.5     & IFLYTEK             &  Not Reported         & \url{https://spark-api.xf-yun.com}   & 2023\\
\hline
\end{tabular}
\end{table*}

\subsection{Evaluation Setup
\label{SUBSEC:Evaluation Setup}}

We designed different experimental stages to answer each research question. 
For \textbf{RQ1}, we identified 300 Java web applications from GitHub using keywords such as \textit{``Java web''}, \textit{``Jobs''}, and \textit{``Books''}. 
These keywords were selected from an online statistical resource\footnote{\url{http://5000best.com/websites/}.} that lists the top 5000 famous websites.
We then cloned these selected web applications, and excluded those without web forms and those that could not run properly in the experimental environment.
We also manually confirmed that the remaining applications did contain web forms, and stored their URLs for the later experiments.
Finally, 30 Java web applications, with 146 web forms, were selected for the experiment.
We manually categorized the 146 web forms into five different categories, based on their functional content, following the methodology outlined in previous literature~\cite{liu2023fill}. 
Using the open coding protocol~\cite{seaman1999qualitative}, we invited three product managers, all of whom had over five years of experience and at least a master's-degree level of education,  to assist with the classification. 
Initially, two product managers independently examined the functionality of each web form and assigned it to a category. 
Their annotations were then iteratively reviewed and consolidated. 
Once there was no disagreement in the classification, the final classification was handed over to a third, highly experienced product manager for double-checking.

\begin{figure*}[!h]
    \centering
    \graphicspath{{Graphs/}}
    \includegraphics[width=0.60\textwidth]{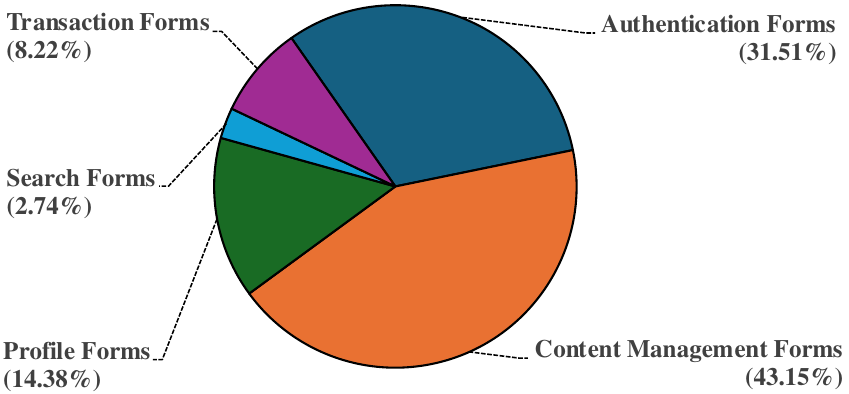}
    \caption{Categories of the web forms under test.}
    \label{FIG: form_categories}
\end{figure*}

The five functional categories of the web forms under test are:
\begin{itemize}
    \item \textit{\textbf{Authentication Forms (31.51\%)}:} 
    Among other things, these forms deal with sign-in, user registration, and password resetting.
    Examples include: 
    Login, Register, and Reset Password forms.

    \item \textit{\textbf{Profile Forms (14.38\%)}:}
    Among other things, these forms allow users to manage and update personal information, account details, or privacy settings.
    Examples include: 
    Edit Personal Information, Add Contact, and Edit Student Information forms.

    \item \textit{\textbf{Content Management Forms (43.15\%)}:}
    Among other things, these forms involve creating, editing, or managing content.
    Examples include: 
    Edit Article, Add Food, and Edit Schedule forms.

    \item \textit{\textbf{Search Forms (2.74\%)}:}
    Among other things, these forms are used for searching and querying data within the application.
    Examples include: 
    Search Artist, Search Reservation, and Search User forms.

    \item \textit{\textbf{Transaction Forms (8.22\%)}:}
    Among other things, these forms handle transactional actions, such as placing orders, making payments, or managing invoices. 
    Examples include: 
    Add Order, Add Payment, and Edit Reservation forms. 
\end{itemize}

We used the {\em successfully-submitted rate} 
(SSR)\footnote{This metric was originally referred to as the ``\textit{form passing rate}''~\cite{alian2024bridging}.} 
to evaluate the effectiveness of the web-form tests:
The SSR is the proportion of LLM-generated web-form tests that could be successfully inserted into the web forms and submitted.
A higher SSR indicates that the generated test cases align better with the expected input and structural requirements of the web forms
---
the LLM has produced valid test cases.

Following previous research~\cite{liu2023fill}, we ran each web form three times, using the 11 LLMs and three types of prompts:
This was a total of $146 \times 11 \times 3 \times 3 = 14,454$ generated web-form tests.
When evaluating the different effectiveness among the 11 LLMs, the two-tailed non-parametric \textit{Mann-Whitney U test} (U-test)~\cite{Wilcoxon1945} and Vargha and Delaney's \textit{effect size} ($\Hat{\textrm{A}}_{12}$)~\cite{Vargha2000} were used:
These have been widely used to detect statistical differences~\cite{Arcuri14,Huang24}.
The U-test uses $p$-values to identify significant differences in SSR results (at a significance level of 5\%):
A $p$-value less than 0.05 indicates a significant difference between the two compared methods~\cite{cowles1982origins,Huang23vpp}.
$\Hat{\textrm{A}}_{12}$ compares two methods, $\mathcal{M}$ and $\mathcal{N}$, as:
\begin{equation}
    \Hat{\textrm{A}}_{12}(\mathcal{M},\mathcal{N}) = (R_1 /X - (X+1)/2)/Y,
\end{equation}
where $R_1$ is the rank sum of the first data group under comparison;
$X$ is the number of observations in the data sample of $\mathcal{M}$; and
$Y$ is the number of observations in the data sample of $\mathcal{N}$.
In general, $\Hat{\textrm{A}}_{12}(\mathcal{M},\mathcal{N})=0.50$ means $\mathcal{M}$ and $\mathcal{N}$ have equal performance;
$\Hat{\textrm{A}}_{12}(\mathcal{M},\mathcal{N})>0.50$ means that $\mathcal{M}$ is better than $\mathcal{N}$; and
$\Hat{\textrm{A}}_{12}(\mathcal{M},\mathcal{N})<0.50$ means that $\mathcal{N}$ is better than $\mathcal{M}$.

For \textbf{RQ2.1}, we conducted a case study to analyze why some generated web-form tests were not successfully submitted.
For \textbf{RQ2.2}, we designed an online questionnaire\footnote{\url{https://github.com/abelli1024/web-form-test-gen-empirical-study}.\label{footnote:study}} to collect user evaluations of the LLM-generated web-form tests.
We invited 20 testers with software testing experience from famous Internet enterprises and research institutions: 
This included five researchers from research institutions (each of whom with a rank of associate professor or higher), and fifteen testers from enterprises (each of whom had over five years of professional experience in software product testing).

For each web form, we presented the screenshot and the corresponding web-form-test text generated by the 11 LLMs and the three types of prompts.
To eliminate any impact of the LLM name, we concealed their names in the questionnaire, only providing the generated web-form-test texts.
We randomly selected 15 web forms from the 146 web forms for each tester.
It should be noted that all the presented web forms were of comparable difficulty, with no significant differences in complexity among them.
We used Kendall’s W~\cite{siegel1957nonparametric} value to measure the agreement among different testers for responses to the questionnaire statements (which used a 5-point Likert scale:
strongly disagree ``1''; 
disagree ``2'';
neutral ``3'';
agree ``4''; and
strongly agree ``5'').
A higher Kendall's W value (close to 1.0) indicates a higher level of agreement among the testers' evaluation results.
We also collected data on the number of testers who use LLMs in their testing processes. 
Additionally, we identified the main concerns of users who currently use or intend to use LLMs for testing.

\subsection{Experiment Environment}
All experiments were conducted on a MacBook Pro laptop with an Apple M3 Max processor and 64GB of RAM. 
The test script was developed in Java.
The version of Google Chrome was 122.0.6261.112 (official version) (arm64), and the version of the Chrome WebDriver was 122.0.6261.128 (r1250580).

\subsection{LLM Access}
In our experiments, the 11 LLMs used for test-case generation were accessed through a combination of paid API access and free access through student membership programs or provided through new user registration offers. 
While these free access conditions allowed us to conduct the experiments without significant costs, we acknowledge that they may not reflect the typical costs associated with the commercial use of these APIs.
This study primarily focused on evaluating the effectiveness of the LLMs for test-case generation:
Our future work will explore the cost-effectiveness of these models in real-world applications.

\section{Results and Discussions
\label{SEC:Experimental Results}}

This section provides the experimental results to answer the research questions.
We also discuss some insights for using LLMs in web-form testing.

\begin{table*}[!t]
\caption{The SSR results on 438 test tasks of three types of prompt design. 
(The symbol $\#$ denotes the successfully-submitted number;
$!\#$ denotes the non-successfully-submitted number; and
$\%$ denotes the SSR.)}
\label{TAB:rq1}
\centering
\scriptsize 
\setlength\tabcolsep{1mm} 
\begin{tabular}{@{}c|l|r|r|r|r|r|r|r|r|r|r|r|l|c@{}}
\hline
\multicolumn{1}{c|}{\multirow{2}{*}{\textbf{No.}}}&\multicolumn{1}{c|}{\multirow{2}{*}{\textbf{Methods}}}
& \multicolumn{3}{c|}{\textbf{RH-P}}  & \multicolumn{3}{c|}{\textbf{LH-P}}  & \multicolumn{3}{c|}{\textbf{PH-P}} & \multicolumn{3}{c|}{\textit{\textbf{Average}}} 
&\multirow{2}{*}{\begin{tabular}[c]{@{}c@{}}\textbf{Best}  \\ \textbf{Performer}\end{tabular}} 
\\\cline{3-14}
 && \multicolumn{1}{r|}{$\#$} & \multicolumn{1}{r|}{$!\#$} & \multicolumn{1}{r|}{$\%$} & \multicolumn{1}{r|}{$\#$} & \multicolumn{1}{r|}{$!\#$} & \multicolumn{1}{r|}{$\%$} & \multicolumn{1}{r|}{$\#$} & \multicolumn{1}{r|}{$!\#$} & \multicolumn{1}{r|}{$\%$} &\multicolumn{1}{r|}{$\#$} & \multicolumn{1}{r|}{$!\#$} & \multicolumn{1}{r|}{$\%$}\\
\hline
1&GPT-3.5 & 198 & 240 & 45.21\% & 299 & 139 & 68.26\% & 365 & 73  & 83.33\% &287.33	&150.67	&65.60\% &PH-P\\
2&GPT-4 & 433 & 5 & \textbf{\textit{98.86\%}} & 425 & 13  & \textbf{\textit{97.03\%}} & 436 & 2 & \textbf{\textit{99.54\%}} & 431.33	&6.67	& \textbf{\textit{98.48\%}} &PH-P\\
3&GLM-3 & 339 & 99  & 77.40\% & 292 & 146 & 66.67\% & 350 & 88  & 79.91\% &327.00	&111.00&	74.66\% &PH-P\\
4&GLM-4 & 378 & 60  & 86.30\% & 399 & 39  & 91.10\% & 399 & 39  & 91.10\% &392.00&	46.00&	89.50\% &LH-P\&PH-P\\
5&GLM-4V  & 0 & 438 & 0.00\%  & 0 & 438 & 0.00\%  & 0 & 438 & 0.00\% &0.00 & 438.00 & 0.00\% &--\\
6&Baichuan2 & 371 & 67  & 84.70\% & 380 & 58  & 86.76\% & 419 & 19  & 95.66\% &390.00	&48.00	&89.04\% &PH-P\\
7&LLaMa2(7B)  & 151 & 287 & 34.47\% & 1 & 437 & 0.23\%  & 185 & 253 & 42.24\% & 112.33	& 325.67	& 25.65\% &PH-P\\
8&LLaMa2(13B) & 188 & 250 & 42.92\% & 4 & 434 & 0.91\%  & 177 & 261 & 40.41\% & 123.00	&315.00	&28.08\% &RH-P\\
9&LLaMa2(70B) & 344 & 94  & 78.54\% & 29  & 409 & 6.62\%  & 338 & 100 & 77.17\% &237.00	&201.00	&54.11\% &RH-P\\
10&Spark-3 & 202 & 236 & 46.12\% & 215 & 223 & 49.09\% & 316 & 122 & 72.15\% &244.33&	193.67	&55.78\% &PH-P\\ 
11&Spark-3.5 & 297 & 141 & 67.81\% & 378 & 60  & 86.30\% & 418 & 20  & 95.43\% & 364.33&	73.67&	83.18\% &PH-P\\
\hline
\multicolumn{2}{c|}{\textit{\textbf{Average}}} & 263.73 & 174.27 & 60.21\% & 220.18 & 217.82 & 50.27\% & 309.36 & 128.64 & \textbf{\textit{70.63\%}} & 264.42&	173.58	&60.37\% &PH-P\\
\hline
\end{tabular}
\end{table*}

\begin{figure*}[!t]
\centering
    \subfigure[RH-P]{\includegraphics[width=0.47\textwidth]{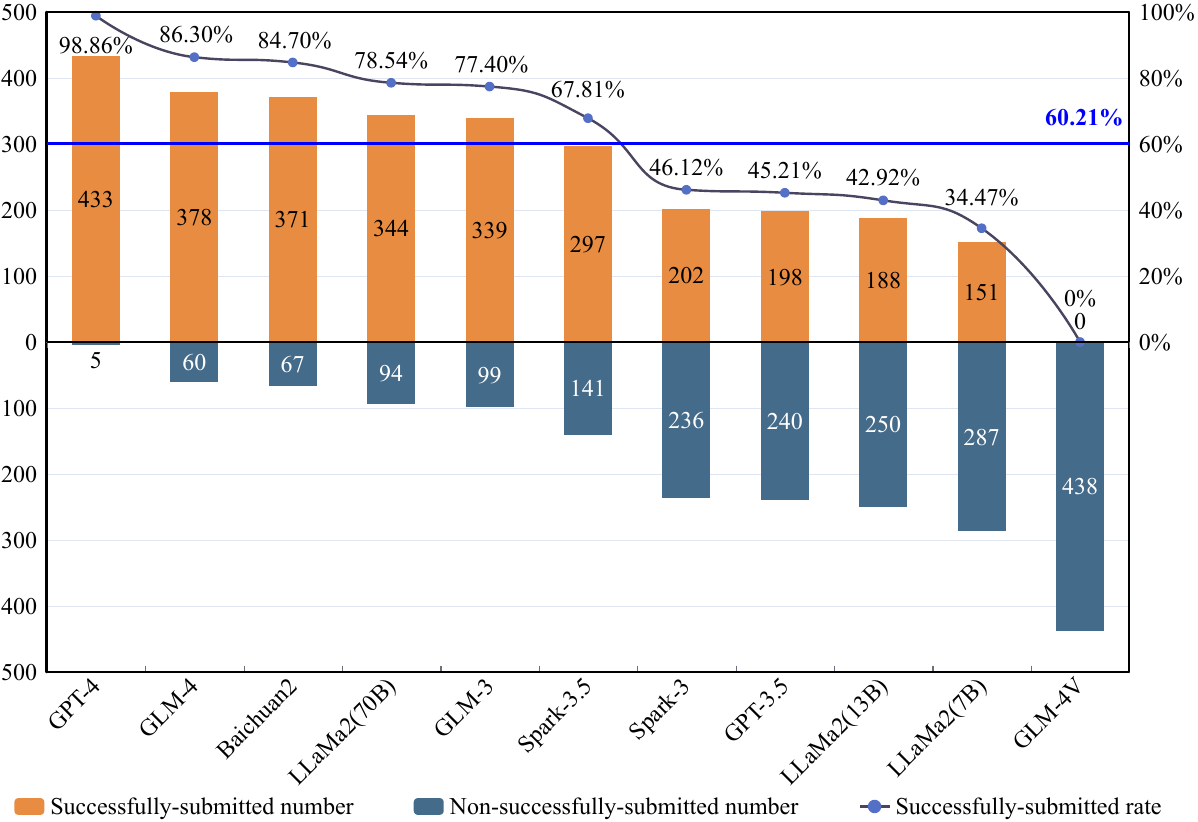}\label{FIG:RH-P effectiveness}}
    \subfigure[LH-P]{\includegraphics[width=0.47\textwidth]{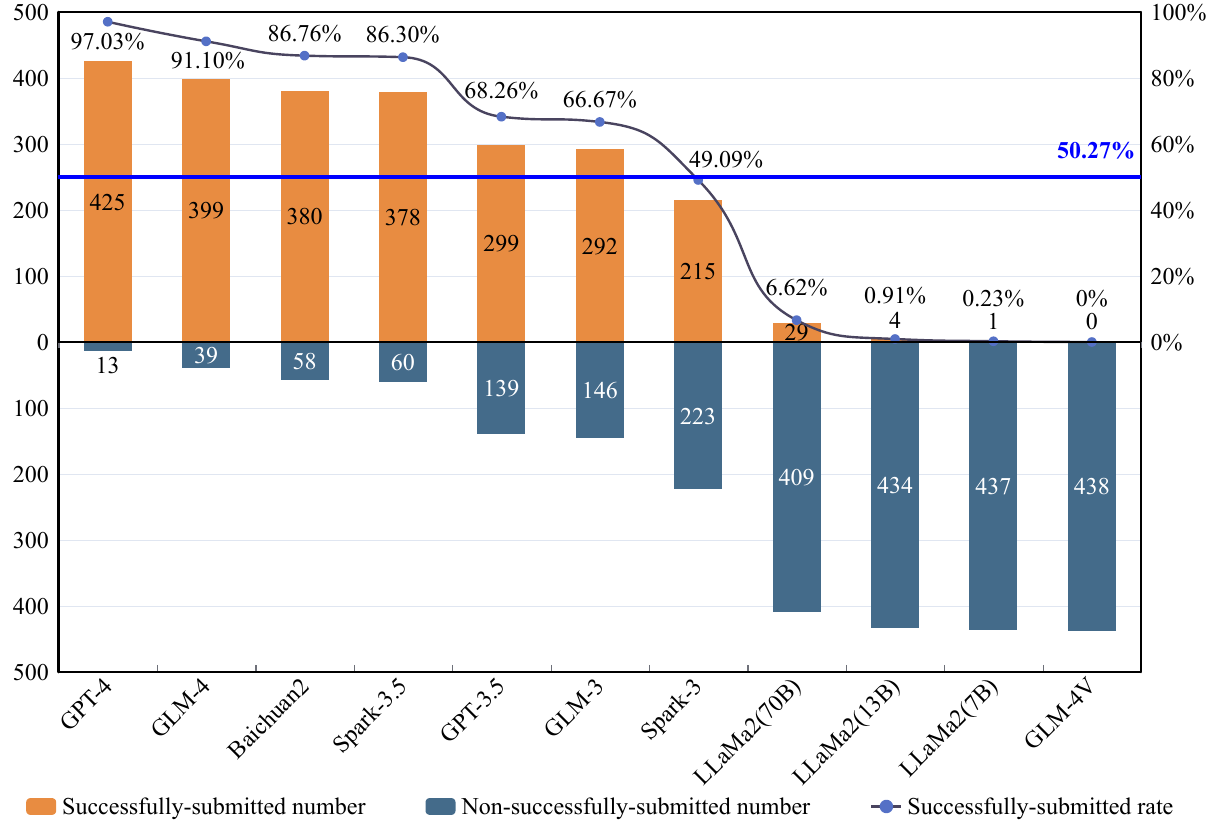}\label{FIG:LH-P effectiveness}}
    
    \subfigure[PH-P]{\includegraphics[width=0.47\textwidth]{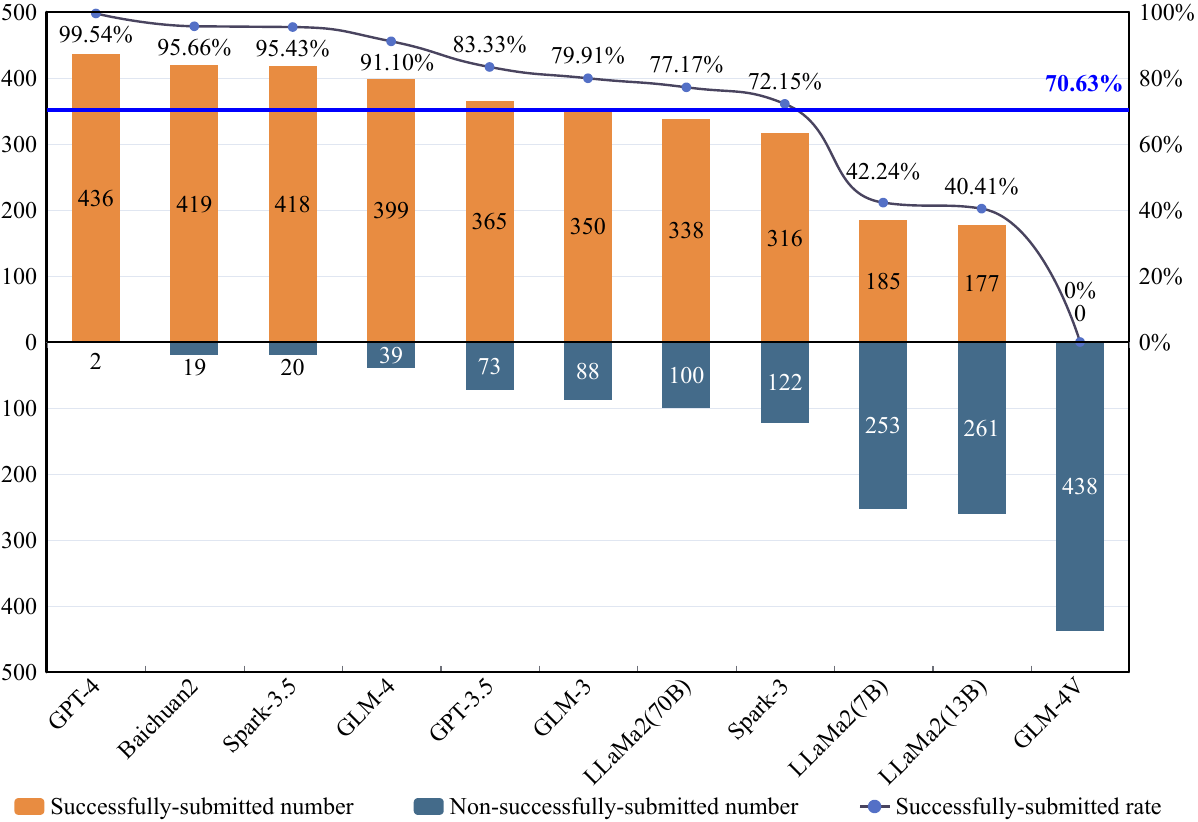}\label{FIG:PH-P effectiveness}}
    \subfigure[\textit{Average}]{\includegraphics[width=0.47\textwidth]{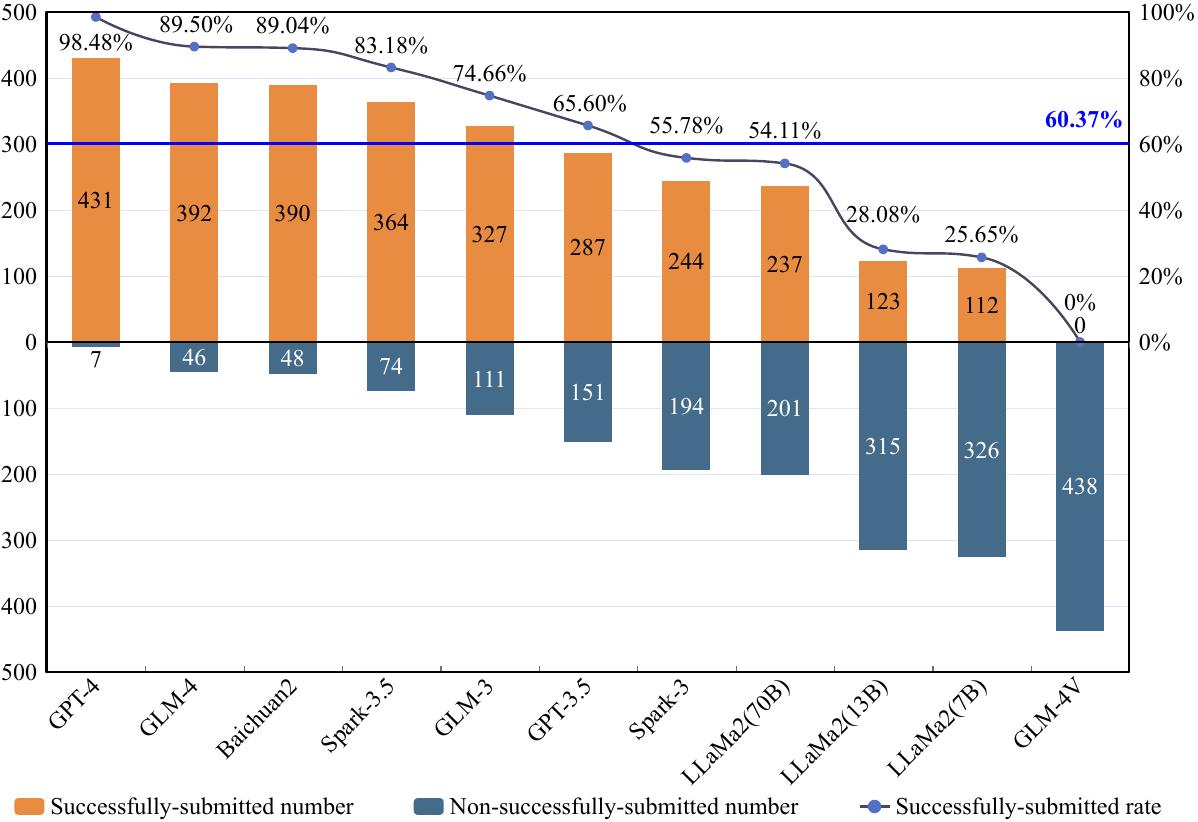}\label{FIG:average effectiveness}}
    
\caption{The sorted SSR results on 438 test tasks of the three types of prompt design.}
\label{FIG: rq1 effectiveness}
\end{figure*}

\subsection{RQ1: How effective are the web-form tests generated by different LLMs?
\label{sec: results-rq1}}
This section discusses the effectiveness of the LLM-generated web-form tests.
We provide an analysis from the perspective of the three prompt types.

\subsubsection{Answer to RQ1.1} 
\label{sec: Answer to RQ1.1}

Table \ref{TAB:rq1} and Figure \ref{FIG: rq1 effectiveness} present the SSR results of the 438 test tasks for the three types of prompts (RH-P, LH-P, and PH-P).
Based on these results, we have the following observations:
\begin{itemize}
    \item The average SSR results for RH-P, LH-P, and PH-P are 60.21\%, 50.27\%, and 70.63\%, respectively, indicating that PH-P can guide LLMs to generate better web-form tests than RH-P or LH-P.

    \item The RH-P SSR ranges from 0.00\% to 98.86\%, the LH-P SSR ranges from 0.00\% to 97.03\%, and the PH-P SSR ranges from 0.00\% to 99.54\%.
    For all three types of prompts, GPT-4 always has the highest SSR results, indicating the best performance.
    GLM-4V always has an SSR of 0.00\%, and may not be suitable for generating appropriate web-form tests.
   
    Apart from GLM-4V, LLaMa2(7B) performed worst with the RH-P and LH-P types, with SSRs of 34.47\% and 0.23\%, respectively; and 
    LLaMa2(13B) performed worst with the PH-P type, with an SSR of 40.41\%.

    \item For RH-P (Figure \ref{FIG:RH-P effectiveness}), the SSR results of six LLMs (GPT-4, GLM-4, Baichuan2, LLaMa2(70B), GLM-3, and Spark-3.5) are greater than the average (60.21\%), the other five LLMs (Spark-3, GPT-3.5, LLaMa2(7B), LLaMa2(13B), and GLM-4V) have poorer performance.

    \item For LH-P (Figure \ref{FIG:LH-P effectiveness}), the SSR results of six LLMs (GPT-4, GLM-4, Baichuan2, Spark-3.5, GPT-3.5, and GLM-3) are greater than the average (50.27\%), the other five LLMs (Spark-3, LLaMa2(7B), LLaMa2(13B), LLaMa2(70B), and GLM-4V) have poorer performance.

    \item For PH-P (Figure \ref{FIG:PH-P effectiveness}), the SSR results of seven LLMs (GPT-4, Baichuan2, Spark-3.5, GLM-4, GPT-3.5, GLM-3, LLaMa2(70B), and Spark-3) are greater than the average (70.63\%), the other three LLMs (LLaMa2(7B), LLaMa2(13B), and GLM-4V) have poorer performance.

    \item 
    The average SSR results for RH-P, LH-P, and PH-P are 60.21\%, 50.27\%, and 70.63\%, respectively.
    Only PH-P achieved a higher average SSR than the overall average result (60.37\%): RH-P and LH-P were both lower than the overall average.
\end{itemize}

Based on the experimental results, we observed that GLM-4V always had an SSR of 0.00\%, regardless of the prompt type. 
This result was unexpected, and requires further analysis.
GLM-4V, originally optimized for vision tasks, was chosen for this study to explore its potential applicability in web-form-test generation. 
While the model excels in processing visual data, it is primarily designed to handle images and visual patterns, not textual or structured data like HTML elements and web forms. 
The results clearly show that GLM-4V was unable to generate appropriate web-form tests (SSR of 0.00\%). 
This suggests that the model's architecture is not suited to process the structured, text-based nature of web forms
--- 
something that requires understanding the interactions among the HTML elements and form behaviors.
Instead of GLM-4V, models that have been specifically trained for natural language processing and text-based tasks, such as GPT-4, should be more appropriate for generating web-form tests.

We next discuss the different performances of the three types of prompts.

\begin{itemize}
    \item 
    With RH-P, we directly feed the raw pruned HTML of the web forms into the LLMs.
    This enables the LLMs to use the contextual information from the HTML to generate the web-form tests. 
    However, we found that some LLMs could not generate effective web-form tests based on the HTML context.
    For example, hint texts in the raw HTML context are descriptions displayed in the web forms that guide users towards what should be entered. 
    However, some LLMs (such as LLaMa2(13B), LLaMa2(70B), and Baichuan2) only returned the hint texts (e.g., ``Please enter the user name''), without generating any web-form tests.

    \item With LH-P and PH-P, we used two approaches to proceed with the pruned HTML: with the LLMs, and with the automated-testing tool.
    We found that the automated testing tool could perform better for the web-form-test generation.
    One of the reasons is that LH-P cannot process the HTML into JSON objects that conform to the expectations: 
    Some key contextual information used for guiding the web-form-test generation may be missed when getting the JSON objects, resulting in the LLMs not successfully generating the tests.
    In contrast, the automated testing tool could process the pruned HTML without missing any key contextual information. 
    Therefore, compared to the automated testing tool, LLMs appear less suitable for parsing the HTML into JSON objects.
\end{itemize}

We determined that the differences in SSR results across the LLMs were due to variations in the model architecture, HTML processing, and context understanding.  
Larger models, with more parameters and diverse training data, demonstrated superior performance by effectively parsing complex HTML structures and capturing relationships among form elements. 
This resulted in higher SSRs across all prompt types. 
In contrast, smaller models, like LLaMa2(7B) and GLM-4V, had trouble with the web-form structures. 
LLaMa2(7B) failed to return the required content, while GLM-4V, optimized for vision tasks, performed poorly across all prompts and is unsuitable for text-based web-form-test generation.

The differences are also related to how each model processed the provided HTML context. 
With RH-P, some models were unable to generate valid tests, often only returning the hint text (e.g., ‘Please enter your username’). 
With LH-P and PH-P, despite providing structured prompts, some models failed to convert the HTML into usable JSON, missing key contextual information. 
In contrast, the automated testing tools were more reliable at parsing and processing the HTML without losing essential details, leading to more effective test-case generation. 
Ultimately, providing more structured and complete context in the prompts (e.g., PH-P) led to higher SSRs, emphasizing the importance of context consistency for effective web-form-test generation.

\begin{tcolorbox}[breakable,colframe=black,colback=white,arc=0mm,left={1mm},top={1mm},bottom={1mm},right={1mm},boxrule={0.25mm}]
\textit{\textbf{Summary of Answers to RQ1.1:}}
Among the three types of prompts, PH-P performed better than RH-P and LH-P. 
Furthermore, the LLMs were found to be less suitable for parsing the HTML of web forms compared to the automated testing tools.
\end{tcolorbox}

\begin{table*}[!t]
\caption{Statistical comparison of the SSR results on 438 test tasks of 11 LLMs with three types of prompt design.}
\label{TAB:statistical analysis}
\scriptsize
\centering
\setlength\tabcolsep{0.9mm}
\begin{tabular}{@{}c|c|ccccccccccc@{}}
\hline
\multirow{3}{*}{\begin{tabular}[c]{@{}c@{}}\textbf{Prompt}  \\ \textbf{types}\end{tabular}} & \multirow{3}{*}{\textbf{Methods}} & \multicolumn{11}{c}{\textbf{Statistical comparison}} \\ \cline{3-13} 
 & & GPT-3.5 & GPT-4 & GLM-3 & GLM-4 & GLM-4V  & Baichuan2 & \begin{tabular}[c]{@{}c@{}}LLaMa2\\(7B)\end{tabular}  & \begin{tabular}[c]{@{}c@{}}LLaMa2\\(13B)\end{tabular} & \begin{tabular}[c]{@{}c@{}}LLaMa2\\(70B)\end{tabular} & Spark-3 & Spark-3.5 \\ \hline
\multirow{11}{*}{RH-P} 
  & GPT-3.5 & -- & \ding{54} (0.23) & \ding{54} (0.34) & \ding{54} (0.29) & \ding{52} (0.73) & \ding{54} (0.30) & \ding{52} (0.55) & \ding{109} (0.51) & \ding{54} (0.33) & \ding{109} (0.50) & \ding{54} (0.39)\\
 & GPT-4 & \ding{52} (0.77) & -- & \ding{52} (0.61) & \ding{52} (0.56) & \ding{52} (0.99) & \ding{52} (0.57) & \ding{52} (0.82) & \ding{52} (0.78) & \ding{52} (0.60) & \ding{52} (0.76) & \ding{52} (0.66) \\
 & GLM-3 & \ding{52} (0.66) & \ding{54} (0.39) & -- & \ding{54} (0.46) & \ding{52} (0.89) & \ding{54} (0.46) & \ding{52} (0.71) & \ding{52} (0.67) & \ding{109} (0.49) & \ding{52} (0.66) & \ding{52} (0.55) \\
 & GLM-4 & \ding{52} (0.71) & \ding{54} (0.44) & \ding{52} (0.54) & -- & \ding{52} (0.93) & \ding{109} (0.51) & \ding{52} (0.76) & \ding{52} (0.72) & \ding{52} (0.54) & \ding{52} (0.70) & \ding{52} (0.59) \\
 & GLM-4V & \ding{54} (0.27) & \ding{54} (0.01) & \ding{54} (0.11) & \ding{54} (0.07) & -- & \ding{54} (0.08) & \ding{54} (0.33) & \ding{54} (0.29) & \ding{54} (0.11) & \ding{54} (0.27) & \ding{54} (0.16) \\
 & Baichuan2 & \ding{52} (0.70) & \ding{54} (0.43) & \ding{52} (0.54) & \ding{109} (0.49) & \ding{52} (0.92) & -- & \ding{52} (0.75) & \ding{52} (0.71) & \ding{52} (0.53) & \ding{52} (0.69) & \ding{52} (0.58) \\
 & LLaMa2(7B) & \ding{54} (0.45) & \ding{54} (0.18) & \ding{54} (0.29) & \ding{54} (0.24) & \ding{52} (0.67) & \ding{54} (0.25) & -- & \ding{54} (0.46) & \ding{54} (0.28) & \ding{54} (0.44) & \ding{54} (0.33) \\
 & LLaMa2(13B) & \ding{109} (0.49) & \ding{54} (0.22) & \ding{54} (0.33) & \ding{54} (0.28) & \ding{52} (0.71) & \ding{54} (0.29) & \ding{52} (0.54) & -- & \ding{54} (0.32) & \ding{109} (0.48) & \ding{54} (0.38) \\
 & LLaMa2(70B) & \ding{52} (0.67) & \ding{54} (0.40) & \ding{109} (0.51) & \ding{54} (0.46) & \ding{52} (0.89) & \ding{54} (0.47) & \ding{52} (0.72) & \ding{52} (0.68) & -- & \ding{52} (0.66) & \ding{52} (0.55) \\
 & Spark-3 & \ding{109} (0.50) & \ding{54} (0.24) & \ding{54} (0.34) & \ding{54} (0.30) & \ding{52} (0.73) & \ding{54} (0.31) & \ding{52} (0.56) & \ding{109} (0.52) & \ding{54} (0.34) & -- & \ding{54} (0.39)\\
 & Spark-3.5 & \ding{52} (0.61) & \ding{54} (0.34) & \ding{54} (0.45) & \ding{54} (0.41) & \ding{52} (0.84) & \ding{54} (0.42) & \ding{52} (0.67) & \ding{52} (0.62) & \ding{54} (0.45) & \ding{52} (0.61) & -- \\ \hline
 
\multirow{11}{*}{LH-P} 
 & GPT-3.5 & -- & \ding{54} (0.36) & \ding{109} (0.51) & \ding{54} (0.39) & \ding{52} (0.84) & \ding{54} (0.41) & \ding{52} (0.84) & \ding{52} (0.84) & \ding{52} (0.81) & \ding{52} (0.60) & \ding{54} (0.41)\\
 & GPT-4 & \ding{52} (0.64) & -- & \ding{52} (0.65) & \ding{52} (0.53) & \ding{52} (0.99) & \ding{52} (0.55) & \ding{52} (0.98) & \ding{52} (0.98) & \ding{52} (0.95) & \ding{52} (0.74) & \ding{52} (0.55)\\
 & GLM-3 & \ding{109} (0.49) & \ding{54} (0.35) & -- & \ding{54} (0.38) & \ding{52} (0.83) & \ding{54} (0.40) & \ding{52} (0.83) & \ding{52} (0.83) & \ding{52} (0.80) & \ding{52} (0.59) & \ding{54} (0.40) \\
 & GLM-4 & \ding{52} (0.61) & \ding{54} (0.47) & \ding{52} (0.62) & -- & \ding{52} (0.96) & \ding{52} (0.52) & \ding{52} (0.95) & \ding{52} (0.95) & \ding{52} (0.92) & \ding{52} (0.71) & \ding{52} (0.52) \\
 & GLM-4V & \ding{54} (0.16) & \ding{54} (0.01) & \ding{54} (0.17) & \ding{54} (0.04) & -- & \ding{54} (0.07) & \ding{109} (0.50) & \ding{54} (0.50) & \ding{54} (0.47) & \ding{54} (0.25) & \ding{54} (0.07) \\
 & Baichuan2 & \ding{52} (0.59) & \ding{54} (0.45) & \ding{52} (0.60) & \ding{54} (0.48) & \ding{52} (0.93) & -- & \ding{52} (0.93) & \ding{52} (0.93) & \ding{52} (0.90) & \ding{52} (0.69) & \ding{109} (0.50) \\
 & LLaMa2(7B) & \ding{54} (0.16) & \ding{54} (0.02) & \ding{54} (0.17) & \ding{54} (0.05) & \ding{109} (0.50) & \ding{54} (0.07) & -- & \ding{109} (0.50) & \ding{54} (0.47) & \ding{54} (0.26) & \ding{54} (0.07) \\
 & LLaMa2(13B) & \ding{54} (0.16) & \ding{54} (0.02) & \ding{54} (0.17) & \ding{54} (0.05) & \ding{52} (0.50) & \ding{54} (0.07) & \ding{109} (0.50) & -- & \ding{54} (0.47) & \ding{54} (0.26) & \ding{54} (0.07) \\
 & LLaMa2(70B) & \ding{54} (0.19) & \ding{54} (0.05) & \ding{54} (0.20) & \ding{54} (0.08) & \ding{52} (0.53) & \ding{54} (0.10) & \ding{52} (0.53) & \ding{52} (0.53) & -- & \ding{54} (0.29) & \ding{54} (0.10) \\
 & Spark-3 & \ding{54} (0.40) & \ding{54} (0.26) & \ding{54} (0.41) & \ding{54} (0.29) & \ding{52} (0.75) & \ding{54} (0.31) & \ding{52} (0.74) & \ding{52} (0.74) & \ding{52} (0.71) & -- & \ding{54} (0.31) \\
 & Spark-3.5 & \ding{52} (0.59) & \ding{54} (0.45) & \ding{52} (0.60) & \ding{54} (0.48) & \ding{52} (0.93) & \ding{109} (0.50) & \ding{52} (0.93) & \ding{52} (0.93) & \ding{52} (0.90) & \ding{52} (0.69) & --\\ \hline
 
\multirow{11}{*}{PH-P} 
 & GPT-3.5 & -- & \ding{54} (0.42) & \ding{109} (0.52) & \ding{54} (0.46) & \ding{52} (0.92) & \ding{54} (0.44) & \ding{52} (0.71) & \ding{52} (0.71) & \ding{52} (0.53) & \ding{52} (0.56) & \ding{54} (0.44)\\
 & GPT-4 & \ding{52} (0.58) & -- & \ding{52} (0.60) & \ding{52} (0.54) & \ding{52} (1.00) & \ding{52} (0.52) & \ding{52} (0.79) & \ding{52} (0.80) & \ding{52} (0.61) & \ding{52} (0.64) & \ding{52} (0.52) \\
 & GLM-3 & \ding{109} (0.48) & \ding{54} (0.40) & -- & \ding{54} (0.44) & \ding{52} (0.90) & \ding{54} (0.42) & \ding{52} (0.69) & \ding{52} (0.70) & \ding{109} (0.51) & \ding{52} (0.54) & \ding{54} (0.42) \\
 & GLM-4 & \ding{52} (0.54) & \ding{54} (0.46) & \ding{52} (0.56) & -- & \ding{52} (0.96) & \ding{54} (0.48) & \ding{52} (0.74) & \ding{52} (0.75) & \ding{52} (0.57) & \ding{52} (0.59) & \ding{54} (0.48) \\
 & GLM-4V & \ding{54} (0.08) & \ding{54} (0.00) & \ding{54} (0.10) & \ding{54} (0.04) & -- & \ding{54} (0.02) & \ding{54} (0.29) & \ding{54} (0.30) & \ding{54} (0.11) & \ding{54} (0.14) & \ding{54} (0.02) \\
 & Baichuan2 & \ding{52} (0.56) & \ding{54} (0.48) & \ding{52} (0.58) & \ding{52} (0.52) & \ding{52} (0.98) & -- & \ding{52} (0.77) & \ding{52} (0.78) & \ding{52} (0.59) & \ding{52} (0.62) & \ding{109} (0.50)\\
 & LLaMa2(7B) & \ding{54} (0.29) & \ding{54} (0.21) & \ding{54} (0.31) & \ding{54} (0.26) & \ding{52} (0.71) & \ding{54} (0.23) & -- & \ding{109} (0.51) & \ding{54} (0.33) & \ding{54} (0.35) & \ding{54} (0.23)\\
 & LLaMa2(13B) & \ding{54} (0.29) & \ding{54} (0.20) & \ding{54} (0.30) & \ding{54} (0.25) & \ding{52} (0.70) & \ding{54} (0.22) & \ding{109} (0.49) & -- & \ding{54} (0.32) & \ding{54} (0.34) & \ding{54} (0.22) \\
 & LLaMa2(70B) & \ding{54} (0.47) & \ding{54} (0.39) & \ding{109} (0.49) & \ding{54} (0.43) & \ding{52} (0.89) & \ding{54} (0.41) & \ding{52} (0.67) & \ding{52} (0.68) & -- & \ding{109} (0.53) & \ding{54} (0.41) \\
 & Spark-3 & \ding{54} (0.44) & \ding{54} (0.36) & \ding{54} (0.46) & \ding{54} (0.41) & \ding{52} (0.86) & \ding{54} (0.38) & \ding{52} (0.65) & \ding{52} (0.66) & \ding{109} (0.47) & -- & \ding{54} (0.38) \\
 & Spark-3.5 & \ding{52} (0.56) & \ding{54} (0.48) & \ding{52} (0.58) & \ding{52} (0.52) & \ding{52} (0.98) & \ding{109} (0.50) & \ding{52} (0.77) & \ding{52} (0.78) & \ding{52} (0.59) & \ding{52} (0.62) & -- \\ \hline
\end{tabular}
\end{table*}

\subsubsection{Answer to RQ1.2
\label{sec: Answer to RQ1.2}}

Table \ref{TAB:statistical analysis} presents statistical comparisons of the SSR results for the 11 LLMs:
For any two LLMs, $\mathcal{M}$ and $\mathcal{N}$,
``\ding{109}'' indicates that there was no statistical difference between them (the $p$-value was greater than 0.05);
``\ding{52}'' means that $\mathcal{M}$ was significantly better ($p$-value was less than 0.05, and the SSR $\Hat{\textrm{A}}_{12}(\mathcal{M},\mathcal{N})$ was greater than 0.50); and
``\ding{54}'' means that $\mathcal{N}$ was significantly better ($p$-value was less than 0.05, and the SSR $\Hat{\textrm{A}}_{12}(\mathcal{M},\mathcal{N})$ was less than 0.50 ).
The numbers in parentheses following ``\ding{52}'', ``\ding{109}'', or ``\ding{54}'' are the $\Hat{\textrm{A}}_{12}$ values of the SSR difference between the two LLMs (for that specific prompt type).
For example, for the RH-P prompt type, the comparison between GPT-3.5 and GPT-4 has a result of \ding{54}(0.23), meaning that the  GPT-3.5 SSR value was 23\% lower than that of GPT-4 with the RH-P prompt type.
Based on the results, we have the following observations:

\begin{itemize} 
    \item 
    Across all three prompt types, GPT-4 significantly outperformed GPT-3.5, with SSR improvements of 77\%, 64\%, and 58\% for RH-P, LH-P, and PH-P, respectively. 
    GPT-4's performance was particularly outstanding with PH-P. 
    In contrast, GLM-4V always performed poorly, with its SSR decreasing by as much as 100\% with PH-P.
    
    \item 
    With RH-P, compared to GPT-3.5, the SSR of GLM-3, GLM-4, Baichuan2, and Spark-3.5 was 34\%, 29\%, 30\%, and 39\% better, respectively, showing strong performance. 
    However, the SSR of LLaMa2(7B), LLaMa2(13B), LLaMa2(70B), and Spark-3 were 55\%, 51\%, 33\%, and 50\% lower, respectively:
    They had a significantly poorer performance than GPT-3.5.
    The SSR of GLM-3 and GLM-4 was 61\% and 56\% lower than that of GPT-4, respectively. 
    The SSR of the LLaMa2 series (7B, 13B, 70B) was 82\%, 78\%, and 60\% lower; 
    while the SSR of Spark-3 and Spark-3.5 was 76\% and 66\% lower. 
    In contrast, Baichuan2's SSR was only 57\% lower than GPT-4, making was relatively good:
    Its performance was only slightly below that of GLM-4.
    
    \item 
    With LH-P, compared to GPT-3.5, the SSR of GLM-4, Baichuan2, and Spark-3.5 was 39\%, 41\%, and 41\% better, respectively. 
    However, the SSR of LLaMa2(7B), LLaMa2(13B), LLaMa2(70B), and Spark-3 was 84\%, 84\%, 81\%, and 60\% lower, respectively, representing a significantly poorer performance. 
    GLM-3 was slightly better, but still 51\% poorer than GPT-3.5.
    GLM-4's SSR was only 53\% poorer than that of GPT-4, again performing relatively well among the weaker models. 
    In contrast, the LLaMa2 series (7B, 13B, 70B) showed the largest performance gaps, with SSRs lower than GPT-4 by 98\%, 98\%, and 95\%, respectively. 
    Among the remaining models, Spark-3, Spark-3.5, GLM-3, and Baichuan2 had SSR results that were 74\%, 55\%, 65\%, and 55\% lower than GPT-4, respectively.

    \item 
    With PH-P, the SSR of GLM-4, Baichuan2, and Spark-3.5 was 46\%, 44\%, and 44\% better than that of GPT-3.5, respectively. 
    However, the SSR of GLM-3, Spark-3, LLaMa2(7B), LLaMa2(13B), and LLaMa2(70B) was 52\%, 56\%, 71\%, 71\%, and 53\% lower, respectively, showing weaker performance. 
    The SSR of GLM-3 and GLM-4 was 60\% and 54\% lower than that of GPT-4, respectively. 
    The SSR of the LLaMa2 series (7B, 13B, 70B) was 79\%, 80\%, and 61\% lower, respectively, while the SSR of Spark-3 and Spark-3.5 was 64\% and 52\% lower. 
    Baichuan2’s SSR was 52\% lower, which (along with Spark-3.5) was relatively strong for the weaker models.
\end{itemize}

Based on the statistical SSR comparisons (Table \ref{TAB:statistical analysis}), we have the following observations:
\begin{itemize}
    \item GPT-4 has the best performance, significantly outperforming all other LLMs in every comparison. 
    
    \item GLM-4, Baichuan2, and Spark-3.5 also exhibit strong performances, with 24, 23, and 20 out of 30 significantly superior situations across the three types of prompts, respectively.
        
    \item GLM-3, GPT-3.5, LLaMa2(70B), and Spark-3 have moderate performances, with 15, 12, 12, and 9 out of 30 significantly superior situations, respectively.
        
    \item LLaMa2(13B), LLaMa2(7B), and GLM-4V have the fewest significantly superior situations, indicating the weakest performances among the evaluated LLMs.
\end{itemize}

Table~\ref{TAB: ssr_result_form_category} presents an analysis of 
the LLMs' performance across the five categories of web forms
---
authentication forms, profile forms, content-management forms, search forms, and transaction forms (Figure~\ref{FIG: form_categories}). 
Based on these results, we have the following observations:

\begin{itemize}
    \item 
    With RH-P, GPT-4 consistently achieved the highest SSR scores, across all five web-form categories:
    It scores ranged from the lowest of 94.44\% (Transaction Forms) to the perfect score of 100\% (Search Forms).
    Notably, GPT-4 outperformed all other models, for every web-form category. 
    Surprisingly, apart from GLM-4V, GPT-3.5 had the worst performance for the Authentication Forms, with an SSR of only 27.54\%:
    This shows that it had difficulty handling this category of web form, compared to other models.
    
    \item 
    With LH-P, GLM-4 had outstanding performance, achieving the highest SSR (91.67\%) for the Search Forms, outperforming GPT-4. 
    Additionally, for Profile Forms, GLM-4 tied with GPT-4, with both scoring a perfect 100\%. 
    Notably, Spark-3.5 also performed well with the Search Forms, achieving an impressive SSR of 91.67\%, tying with GLM-4 for the best performance. 
    In contrast, LLaMa2(7B) and LLaMa2(13B) performed consistently poorly across all five web form categories, with significantly lower SSRs.
        
    \item 
    With PH-P, Baichuan2, GLM-4, LLaMa2(70B), Spark-3.5, and GPT-4 all achieved a perfect SSR of 100\% with the Search Forms. 
    Similarly, with Transaction Forms, Spark-3.5 was a top performer, tying with GPT-4 with a perfect 100\%.
    In the Content Management Forms, Spark-3.5 came second to GPT-4, with an SSR of 94.71\%. 
    Baichuan2 had the second-highest SSRs for the Authentication Forms (98.55\%), Profile Forms (98.41\%), and Transaction Forms (88.89\%), following GPT-4. 
    Surprisingly, Spark-3 had the second-best SSR of 88.89\% with the Transaction Forms.
\end{itemize}

\begin{table}[!t]
\centering
\scriptsize
\caption{SSR results for 438 test tasks, using three types of prompt design, across five web-form categories. 
(Aut., Con., Pro., Sea., and Tra. denotes the Authentication Forms,
Content Management Forms,
Profile Forms,
Search Forms, and
Transaction Forms, respectively.)}
\label{TAB: ssr_result_form_category}
\setlength\tabcolsep{0.15mm}
\begin{tabular}{c|c|r|r|r|r|r|r|r|r|r|r|r|r|r|r|r}
\hline
\multicolumn{1}{c|}{\multirow{2}{*}{No.}} & \multicolumn{1}{c|}{\multirow{2}{*}{Methods}} & \multicolumn{5}{c|}{RH-P} & \multicolumn{5}{c|}{LH-P} & \multicolumn{5}{c}{PH-P} \\ \cline{3-17}
& & Aut.& Con.& Pro.& Sea. & Tra.& Aut.& Con.& Pro. & Sea.& Tra.& Aut.& Con.& Pro. & Sea. & Tra. \\ \hline
1& GPT-3.5  & 27.54\% & 55.56\% & 46.03\% & 50.00\%& 55.56\% & 67.39\% & 65.08\% & 71.43\%& 58.33\% & 86.11\% & 84.78\% & 82.54\% & 87.30\%& 75.00\%& 77.78\%\\
2& GPT-4& \textit{\textbf{99.28\%}} & \textit{\textbf{99.47\%}} & \textit{\textbf{98.41\%}} & \textit{\textbf{100.00\%}} & \textit{\textbf{94.44\%}} & \textit{\textbf{97.83\%}} & \textit{\textbf{96.30\%}} & \textit{\textbf{100.00\%}} & 83.33\% & \textit{\textbf{97.22\%}} & \textit{\textbf{99.28\%}} & \textit{\textbf{99.47\%}} & \textit{\textbf{100.00\%}} & \textit{\textbf{100.00\%}} & \textit{\textbf{100.00\%}} \\
3& GLM-3& 65.94\% & 80.95\% & 93.65\% & 75.00\%& 75.00\% & 67.39\% & 66.14\% & 65.08\%& 75.00\% & 66.67\% & 87.68\% & 72.49\% & 87.30\%& 75.00\%& 77.78\%\\
4& GLM-4& 90.58\% & 78.84\% & 96.83\% & 91.67\%& 88.89\% & 94.20\% & 88.89\% & \textit{\textbf{100.00\%}} & \textit{\textbf{91.67\%}} & 75.00\% & 97.10\% & 87.83\% & 88.89\%& \textit{\textbf{100.00\%}} & 86.11\%\\
5& GLM-4V & 0.00\%& 0.00\%& 0.00\%& 0.00\% & 0.00\%& 0.00\%& 0.00\%& 0.00\% & 0.00\%& 0.00\%& 0.00\%& 0.00\%& 0.00\% & 0.00\% & 0.00\% \\
6& Baichuan2& 79.71\% & 87.30\% & 85.71\% & 75.00\%& 91.67\% & 87.68\% & 84.13\% & 98.41\%& 41.67\% & 91.67\% & 98.55\% & 93.65\% & 98.41\%& \textit{\textbf{100.00\%}} & 88.89\%\\
7& LLaMa2(7B) & 41.30\% & 33.33\% & 33.33\% & 0.00\% & 27.78\% & 0.00\%& 0.53\%& 0.00\% & 0.00\%& 0.00\%& 50.00\% & 34.92\% & 46.03\%& 41.67\%& 44.44\%\\
8& LLaMa2(13B)& 44.93\% & 38.10\% & 52.38\% & 58.33\%& 38.89\% & 0.00\%& 2.12\%& 0.00\% & 0.00\%& 0.00\%& 46.38\% & 38.62\% & 44.44\%& 25.00\%& 25.00\%\\
9& LLaMa2(70B)& 78.99\% & 73.02\% & 95.24\% & 66.67\%& 80.56\% & 0.00\%& 11.64\% & 0.00\% & 58.33\% & 0.00\%& 81.16\% & 71.96\% & 77.78\%& \textit{\textbf{100.00\%}} & 80.56\%\\
10 & Spark-3& 38.41\% & 42.86\% & 61.90\% & 33.33\%& 69.44\% & 73.19\% & 38.10\% & 44.44\%& 25.00\% & 30.56\% & 65.94\% & 71.43\% & 79.37\%& 66.67\%& 88.89\%\\
11 & Spark-3.5 & 75.36\% & 66.14\% & 61.90\% & 50.00\%& 63.89\% & 86.96\% & 88.36\% & 84.13\%& \textit{\textbf{91.67\%}} & 75.00\% & 95.65\% & 94.71\% & 93.65\%& \textit{\textbf{100.00\%}} & \textit{\textbf{100.00\%}}\\\hline
\end{tabular}
\end{table}

\begin{tcolorbox}
    [breakable,colframe=black,colback=white,arc=0mm,left={1mm},top={1mm},bottom={1mm},right={1mm},boxrule={0.25mm}]
    \textit{\textbf{Summary of Answers to RQ1.2:}}
    
    (1) Some LLMs (such as GPT-4, GLM-4, and Baichuan2) can generate relatively effective web-form tests.
    Among the 11 LLMs, GPT-4 always has the best effectiveness, with the highest SSRs and the best statistical comparison results;
    and 
    GLM-4V always performs worst (SSRs of 0.00\%, and the worst statistical comparison results). 
    Compared with GPT-4, the other LLMs have difficulty generating appropriate tests for web forms, with SSR results between 52\% and 98\% lower.
    Nevertheless, some LLMs (such as GLM-4, Baichuan2, and Spark-3.5) achieve higher SSRs than GPT-3.5, indicating better effectiveness.
    
    (2) The experimental results indicate that both prompt design and LLM selection significantly influence the success rate of the LLM-generated web-form tests.
    Among the three prompt types, PH-P had the highest SSR (70.63\%), outperforming both RH-P (60.21\%) and LH-P (50.27\%):
    This confirms that structured HTML parsing enhances LLM comprehension of web-form context.
    In contrast, RH-P's and LH-P's poorer performance may be explained by
    RH-P retaining excessive HTML details, and LH-P losing key contextual information during processing. 
    These results highlight PH-P's advantage in providing structured input for more effective web-form-test generation.
    GPT-4 (98.48\%) performed best, followed by GLM-4 (89.50\%) and Baichuan2 (89.04\%); while
    GLM-4V (0.00\%) and the LLaMa2 models (25.65\%–54.11\%) had limited effectiveness. 
    These results emphasize the need for high-performing LLMs and structured prompts for robust web-form testing.
    
    (3) With RH-P, GPT-4 had the best performance across different web-form categories. 
    LH-P, GLM-4, and Spark-3.5 stood out for the Search Forms, with an SSR of 91.67\%, outperforming GPT-4. 
    With PH-P, several models tied with GPT-4 for the best SSR (a perfect 100\%) for the Search and Transaction Forms. 
\end{tcolorbox}

\subsection{RQ2: What is the quality of the generated web-form tests?}
This section discusses the quality of the web-form tests generated by the different LLMs.
The section includes an analysis of the reasons why some tests cannot be submitted successfully.
We also include an analysis of the test quality, from a tester’s perspective.

\begin{figure*}[!t]
\centering
    \subfigure[A login web form.]{\includegraphics[width=0.315\textwidth, frame]{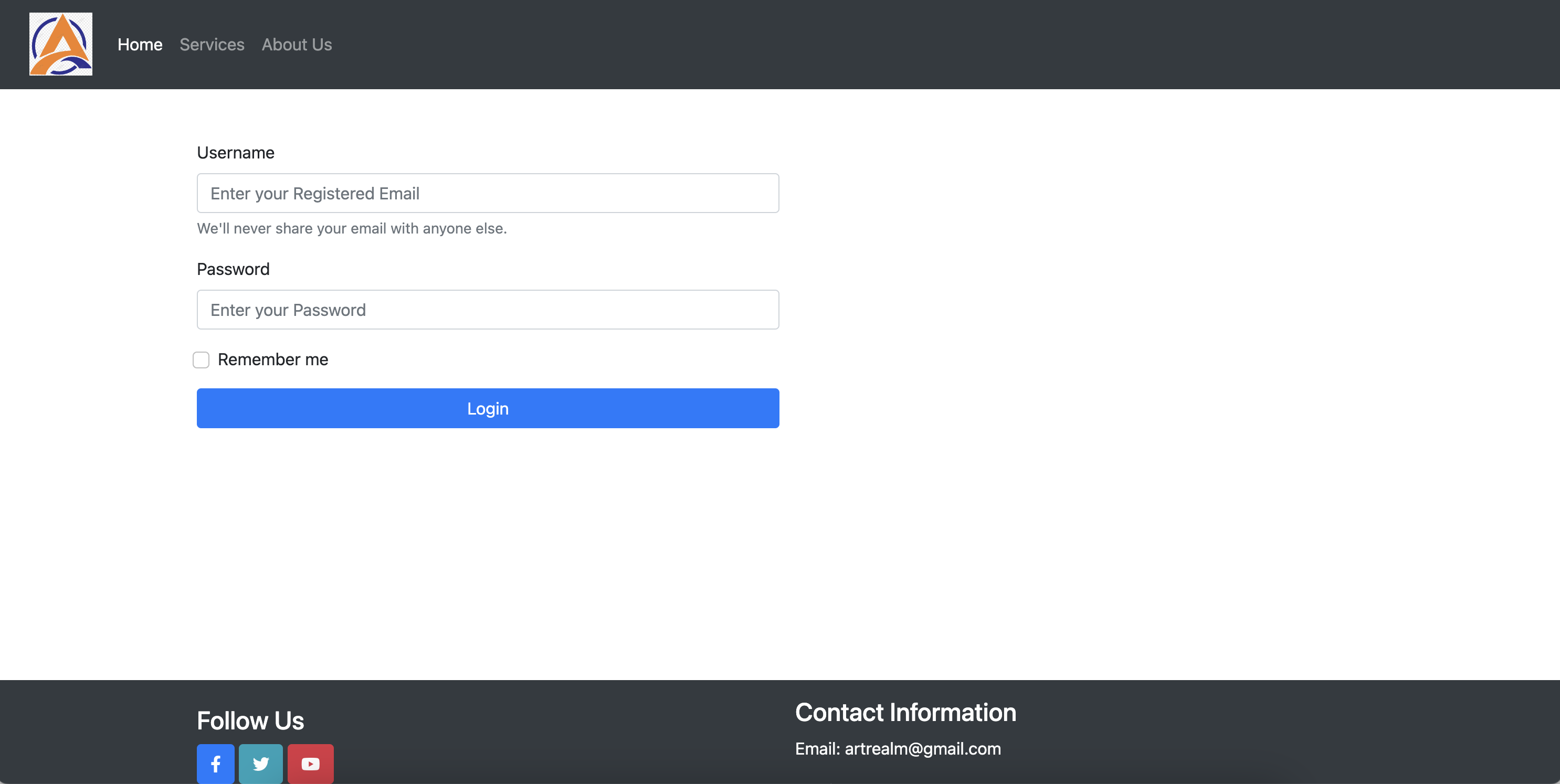}\label{AFIG:example01}}
    \subfigure[A restaurant-reservation web form.]{\includegraphics[width=0.32\textwidth, frame]{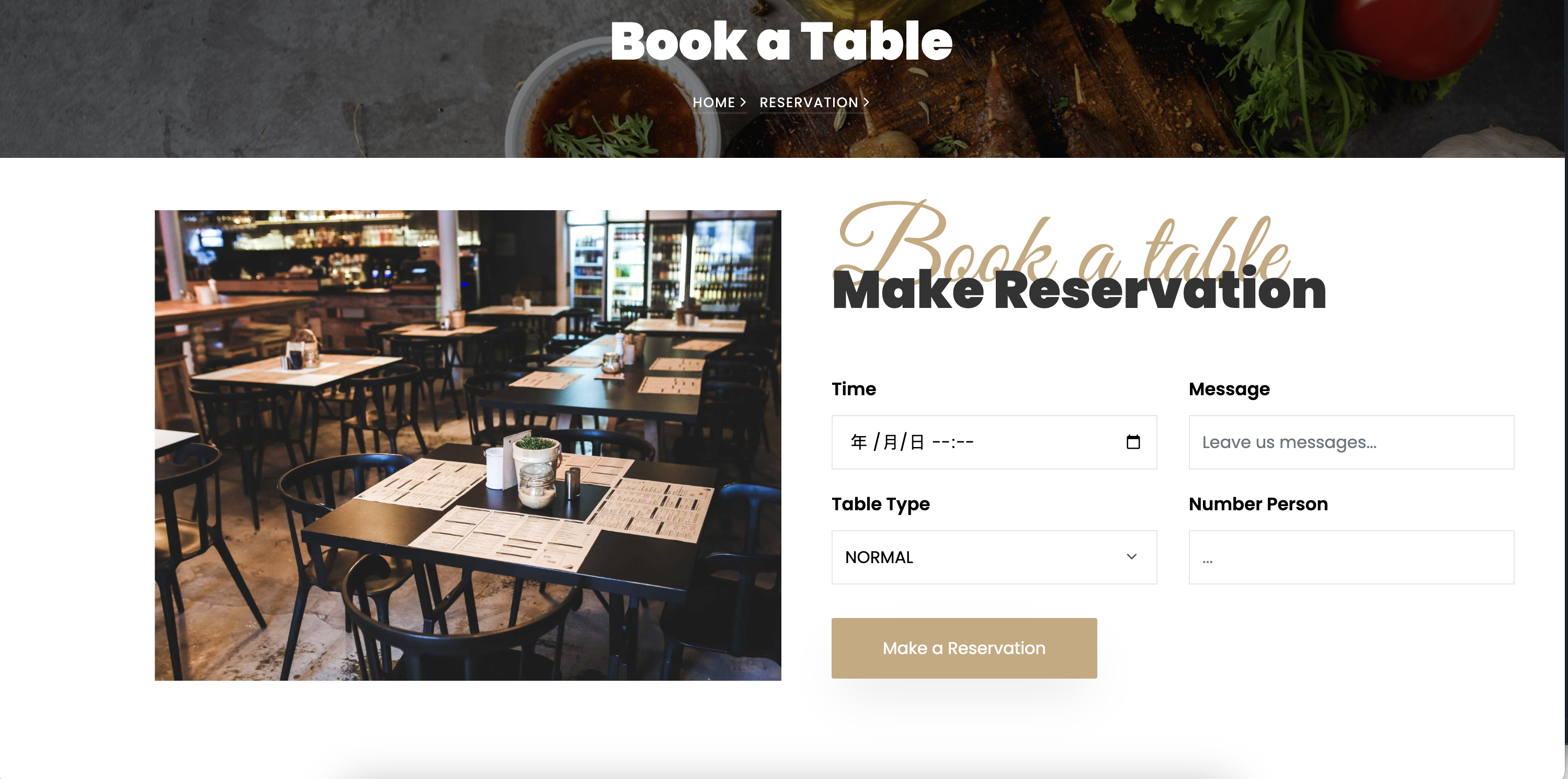}\label{AFIG:example02}}  
    \subfigure[An airplane-booking web form.]{\includegraphics[width=0.32\textwidth, frame]{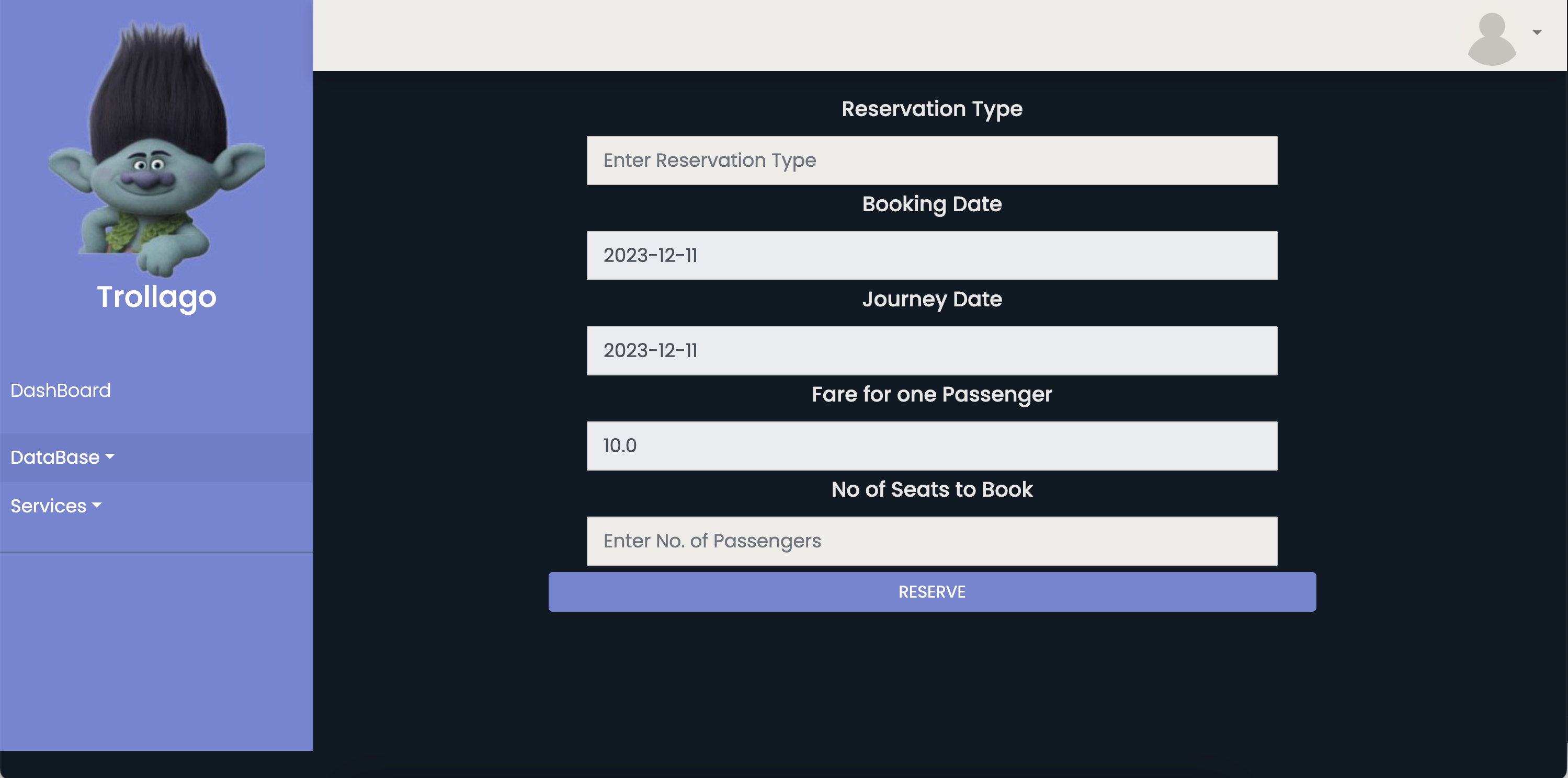}\label{AFIG:example03}}
\caption{Examples of web forms that were not successfully submitted.}
\label{AFIG: examples}
\end{figure*}

\begin{table*}[!t]
\centering
\footnotesize
\setlength\tabcolsep{1.5mm} 
\caption{LLMs and corresponding generated web-form tests for the example in Figure \ref{AFIG:example01}.}
\label{ATAB:example01}
    \begin{tabular}{@{}c|c|p{9.5cm}|l@{}}
    \hline
    \textbf{No.} & \textbf{Methods}     & \textbf{Generated web-form tests}  & \textbf{Status} \\ \hline
    \multirow{5}{*}{1}   & \multirow{5}{*}{GPT-3.5}     & json\{``label'': [``Username'', ``Password'', ``Rememberme''], ``inputType'':[``text'', ``password'', ``checkbox''], ``inputId'': [``username'', ``password'', ``exampleCheck1''], ``inputName'': [``emailId'', ``password''], ``inputValue'': [], ``placeholder'': [], ``additional'': [``We'll never share your email with anyone else.''], ``displayStyle'': [], ``elementType'': [], ``url'': [], ``required'': []\} &  \multirow{5}{*}{\textcolor{red}{\ding{115}} (Reason 1)}\\\hline
    \multirow{2}{*}{2}   & \multirow{2}{*}{GPT-4}       & \#username=john.doe@example.com, \#password=StrongPassword123!, \#exampleChe ck1=true &  \multirow{2}{*}{\textcolor{green}{\ding{108}}}\\\hline
    3   & GLM-3       & \#username=JohnDoe, \#password=Secret123', \#exampleCheck1=true  & \textcolor{green}{\ding{108}} \\\hline
    \multirow{2}{*}{4}   & \multirow{2}{*}{GLM-4}       & \#username=email@example.com, \#password=yourStrong!Passw0rd, \#exampleChe ck1=true & \multirow{2}{*}{\textcolor{green}{\ding{108}}} \\\hline
    \multirow{3}{*}{5}   & \multirow{3}{*}{GLM-4V}      & json\{``selectors'': [``\#username'', ``\#password'', ``\#exampleCheck1'', ``\#submit''], ``values'': \{``\#username'': ``Enter your username here'', ``\#password'': ``Enter your password here'', ``\#exampleCheck1'': ``''\}\}  & \multirow{3}{*}{\textcolor{red}{\ding{115}} (Reason 1)}  \\\hline
    \multirow{2}{*}{6}   & \multirow{2}{*}{Baichuan2}   & \#username=Your Email Address, \#password=Your Password, \#exampleCheck1=Re-member Me &  \multirow{2}{*}{\textcolor{red}{\ding{115}} (Reason 2)} \\\hline
    7   & LLaMa2(7B)  & ``'', ``'', ``'' & \textcolor{red}{\ding{115}} (Reason 3)\\\hline
    \multirow{4}{*}{8}   & \multirow{4}{*}{LLaMa2(13B)} & [\{``label'': ``Username'', ``inputType'': ``text'', ``inputId'': ``username'', ``inputName'': ``emailId'', ``placeholder'': ``Enter your email address'', ``additional'': ``We'll never share your email with anyone else.'', ``displayStyle'': ``block''\}, \{``label'': ``Password'', ``inputType'': ``password  & \multirow{4}{*}{\textcolor{red}{\ding{115}} (Reason 1)}  \\\hline
    \multirow{3}{*}{9}   & \multirow{3}{*}{LLaMa2(70B)} & [\{``label'': ``Username'', ``inputType'': ``text'', ``inputId'': ``username'', ``inputName'': ``emailId'', ``inputValue'': ``'', ``placeholder'': ``'', ``additional'': \{\}, ``displayStyle'': ``block'', ``elementType'': ``div'', ``url'': ``'', ``required'':true\}, \{``label'': ``Password'', ``inputType'': ``'' &  \multirow{3}{*}{\textcolor{red}{\ding{115}} (Reason 1)} \\\hline
    \multirow{2}{*}{10}  & \multirow{2}{*}{Spark-3}     & javascript[`input[id=username]=' + ` ``Username`` `, `input[id=password]=' + ` ``Password'' ', `input[id=exampleCheck1]' + `=true'] & \multirow{2}{*}{\textcolor{red}{\ding{115}} (Reason 1)}  \\\hline
    11 & Spark-3.5   & \#username=emailId, \#password=password, input[id=exampleCheck1]=true  & \textcolor{green}{\ding{108}} \\  \hline      
    \end{tabular}
\end{table*}

\begin{table*}[!t]
\centering
\footnotesize
\setlength\tabcolsep{1.5mm} 
\caption{LLMs and corresponding generated web-form tests for the example in Figure \ref{AFIG:example02}.}
\label{ATAB:example02}
    \begin{tabular}{@{}c|c|p{9.5cm}|l@{}}
    \hline
    \textbf{No.} & \textbf{Methods}     & \textbf{Generated web-form tests}  & \textbf{Status} \\ \hline
    \multirow{2}{*}{1}   & \multirow{2}{*}{GPT-3.5}     & \#time=datetime-local, \#message=text, select\#tableType[name=``tableType''], \#person=number  &  \multirow{2}{*}{\textcolor{red}{\ding{115}} (Reason 2)}\\\hline
    \multirow{2}{*}{2}   & \multirow{2}{*}{GPT-4}       & \#time=2022-12-25T18:30, \#message=Looking forward to the meal, select[name= tableType]=VIP, \#person=4
     &  \multirow{2}{*}{\textcolor{green}{\ding{108}}}\\\hline
    \multirow{2}{*}{3}   & \multirow{2}{*}{GLM-3}       & \#time=Time, \#message=Message, select[name=tableType]={``NORMAL'': ``NORMAL'',``VIP'': ``VIP''}, \#person=Number Person
    & \multirow{2}{*}{\textcolor{red}{\ding{115}} (Reason 2)} \\\hline
    \multirow{2}{*}{4}   & \multirow{2}{*}{GLM-4}       & \#time=2023-04-01T18:30', \#message=Please ensure our table has a great view, \#tableType=VIP, input[name=person]=6
    & \multirow{2}{*}{\textcolor{green}{\ding{108}}} \\\hline
    \multirow{3}{*}{5}   & \multirow{3}{*}{GLM-4V}      & json\{``selectors'': \{``time'':``input[type=`datetime-local']'', ``message'': ``input[type= `text']'', ``tableType'': ``select[name= `tableType']'', ``person'': ``input[type= `number']''\}, ``values'': \{``time'': ``'', ``message'': ``'', ``tableType'': ``NORMAL'', ``person'': ``''\}\}  & \multirow{3}{*}{\textcolor{red}{\ding{115}} (Reason 1)}  \\\hline
    \multirow{2}{*}{6}   & \multirow{2}{*}{Baichuan2}   & \#time=Select time, \#message=Enter message, \#tableType=Choose table type, \#person=Enter number of people  &  \multirow{2}{*}{\textcolor{red}{\ding{115}} (Reason 2)} \\\hline
    \multirow{2}{*}{7}   & \multirow{2}{*}{LLaMa2(7B)}  & \#time=datetime-local();, \#message=text();, .person input[type=number();\}=person();, .tableType select[name=tableType();)=tableType();  & \multirow{2}{*}{\textcolor{red}{\ding{115}} (Reason 2)}  \\\hline
    8   & LLaMa2(13B) & \#time=datetime-local, \#message=text, \#tableType=select, \#person=number  & \textcolor{red}{\ding{115}} (Reason 2)  \\\hline
    9   & LLaMa2(70B) & \#time=datetime-local, \#message=text, \#tableType=select, \#person=number  &  \textcolor{red}{\ding{115}} (Reason 2) \\\hline
    10  & Spark-3     & ``'', ``'', ``'' & \textcolor{red}{\ding{115}} (Reason 3)  \\\hline
    \multirow{2}{*}{11}  & \multirow{2}{*}{Spark-3.5}   & \#time=2023-09-15T18:30', \#message=Hello, I would like to book a table., \#tableType=VIP, \#person=4 & \multirow{2}{*}{\textcolor{green}{\ding{108}} }\\  \hline
    \end{tabular}
\end{table*}

\begin{table*}[!t]
\centering
\footnotesize
\setlength\tabcolsep{1.5mm} 
\caption{LLMs and corresponding generated web-form tests for the example in Figure \ref{AFIG:example03}.}
\label{ATAB:example03}
    \begin{tabular}{@{}c|c|p{9.5cm}|l@{}}
    \hline
    \textbf{No.} & \textbf{Methods}     & \textbf{Generated web-form tests}  & \textbf{Status} \\ \hline
    \multirow{2}{*}{1}   & \multirow{2}{*}{GPT-3.5}     & 
    \#reservationType=sampleReservationType, 
    \#booking\_Date=2023-12-11, 
    \#journeyDate=2023-12-11, 
    \#fare= 10.0,  
    \#noSeats=2  
    &  \multirow{2}{*}{\textcolor{green}{\ding{108}}} \\\hline
    \multirow{2}{*}{2}   & \multirow{2}{*}{GPT-4}       & \#reservationType=Round Trip, \#booking\_Date=2023-12-11, \#journeyDate=2023-12-11, \#fare=10.0, \#noSeats=2  &  \multirow{2}{*}{\textcolor{green}{\ding{108}}}\\\hline
    \multirow{2}{*}{3}   & \multirow{2}{*}{GLM-3}       & \#reservationType=Reservation Type, input[name=scheduleID]=1, \#booking\_Date= 2023-12-11, \#journeyDate=2023-12-11, \#fare=10.0, input[name=noSeats]  & \multirow{2}{*}{\textcolor{red}{\ding{115}} (Reason 2)} \\\hline
    \multirow{3}{*}{4}   & \multirow{3}{*}{GLM-4}       & [`input[name=addReservationForm]', `form'], [`input\#reservationType', `Economy'], [`input\#booking\_Date', `2023-12-11'], [`input\#journeyDate', `2023-12-11'], [`input\#fare', `10.0'], [`input\#noSeats', `2']
    & \multirow{3}{*}{\textcolor{red}{\ding{115}} (Reason 1)} \\\hline
    \multirow{4}{*}{5}   & \multirow{4}{*}{GLM-4V}      & json\{``selectors'': \{``\#reservationType'': ``value from reservationType'', ``\#booking \_Date'': ``value from bookingDate'', ``\#journeyDate'': ``value from journeyDate'', ``\#fare'': ``value from fare'', ``\#noSeats'': ``value from noSeats''\}, ``values'': \{
    ``reservationType'':``'', ``bookingDate'':``2023-12-11'', ``journeyDate'':``2023-12-11'', ``fare'':``10.0'', ``noSeats'':``''\}\}  & \multirow{4}{*}{\textcolor{red}{\ding{115}} (Reason 1)} \\\hline
    \multirow{2}{*}{6}   & \multirow{2}{*}{Baichuan2}   & \#reservationType=Economy, \#booking\_Date=2023-12-11, \#journeyDate=2023-12-11, \#fare=10.0, \#noSeats=10  &  \multirow{2}{*}{\textcolor{green}{\ding{108}}}  \\\hline
    \multirow{3}{*}{7}   & \multirow{3}{*}{LLaMa2(7B)}  & css[\#reservationType=+.\$(`\#reservationType').val()., \#booking\_Date=+.\$(`\#bookingDate') .val()., \#journeyDate=.\$(`\#journeyDate').val()., \#fare=+.\$(`\#fare').val()., \#noSeats=.\$(` \#noSeats').val().]  & \multirow{3}{*}{\textcolor{red}{\ding{115}} (Reason 1)}  \\\hline
    \multirow{4}{*}{8}   & \multirow{4}{*}{LLaMa2(13B)} & 
    \$(`input[name=``reservationType'']').val(``Select a reservation type''); \$(`input[name= ``bookingDate'']').val(``Enter booking date''); 
    \$(`input[name= ``journeyDate'']').val(``Enter journey date''); 
    \$(`input[name= ``fare'']').val(``Enter fare for one passenger''); 
    \$(`input[name= ``noSeats'']').val(``Enter number of seats to book''); 
    & \multirow{4}{*}{\textcolor{red}{\ding{115}} (Reason 1)}  \\\hline
    \multirow{2}{*}{9}   & \multirow{2}{*}{LLaMa2(70B)} & \#reservationType=roundTrip, \#booking\_Date=2023-12-11, \#journeyDate=2023-12-11, \#fare=10.0, \#noSeats=1  &  \multirow{2}{*}{\textcolor{green}{\ding{108}}}  \\\hline
    \multirow{3}{*}{10} & \multirow{3}{*}{Spark-3}     & javascript[input[name=reservationType]=\{reservationType\}, input[name=bookingDate]=\{booking\_Date\}, input[name=journeyDate]=\{journeyDate\}, input[name=fare]=\{fare\}, input[name=noSeats]=\{noSeats\}]  & \multirow{3}{*}{\textcolor{red}{\ding{115}} (Reason 1)}  \\\hline
    \multirow{2}{*}{11}  & \multirow{2}{*}{Spark-3.5}   & \#reservationType=Reservation, \#booking\_Date=2023-12-11, \#journeyDate=2023-12-11, \#fare=10.0, \#noSeats=1  & \multirow{2}{*}{\textcolor{green}{\ding{108}}} \\  \hline
    \end{tabular}
\end{table*}

\subsubsection{Answer to RQ2.1}

Figure \ref{AFIG: examples} presents three web-form instances. 
Tables \ref{ATAB:example01} to \ref{ATAB:example03} show the web-form test information generated with these three web forms by the 11 LLMs.
A ``\textcolor{green}{\ding{108}}'' indicates that the context of the web-form test was consistent with the web form, and could lead to successful submission;
``\textcolor{red}{\ding{115}}'' indicates that the generated web-form test did not satisfy the context, and was not successfully submitted through the web-form.
After analyzing all failed submission situations, we identified three reasons why the LLM-generated web-form tests could not be submitted successfully:

\begin{itemize}
    \item 
    \textbf{Reason 1 (40.82\%):} Some LLMs (such as GLM-3, GPT-3.5, and LLaMa2(7B)) were unable to generate web-form tests in the specific format required by the designed prompts.
    For example, to test the airplane-booking web form (Figure \ref{AFIG:example03}), we used some restrictive prompts when generating web-form tests (Section \ref{SEC:Prompt Design}).
    However, because of the non-determinism of LLMs~\cite{Ouyang23}, some LLMs did not return the correct information, which prevented their correct parsing.
    This was especially the case with LH-P, for which the LLMs first parsed the HTML of a given web-form into a list of JSON structures (Algorithm~\ref{ALG:LLM-Processed HTML}):
    If the LLM could not return the correctly-parsed JSON structures, the subsequent testing process would fail.
    LLMs are constrained by their training datasets, parameter sizes, and other factors, resulting in significant differences in their effectiveness when generating web-form tests.
    Accordingly, some models may fail to generate web-form tests simply because they are unable to understand the specific requirements and restrictions of the prompt.

    \item 
    \textbf{Reason 2 (29.50\%):} Some LLMs (such as GLM-4V, Spark-3, and LLaMa2(13B)) were unable to generate the correct web-form test content based on the provided contextual information.
    For example, when making a table reservation, as shown in Figure \ref{AFIG:example02}, four inputs are needed: 
    dining time, message, table type, and number of diners. 
    Some LLMs (e.g., GLM-3) generated tests with incorrect formatting, such as an inappropriate input for the number of diners. 
    This may be due to some LLMs not correctly parsing the contextual information. 
    
    \item 
    \textbf{Reason 3 (29.68\%):} Connection problems between the testing environment and the LLM API may also have caused the failure or interruption of the web-form-test generation process. 
    With the web form in Figure \ref{AFIG:example01}, for example, LLaMa2(7B) encountered a request timeout when generating the login information:
    It was therefore unable to parse and submit the web-form-test information.
    It should be noted that Reason 3 is not related to the web-form construction, but rather to the communication with LLM APIs:
    The same web form was successfully parsed and processed by other LLMs, without any issues.
\end{itemize}

\begin{tcolorbox}[breakable,colframe=black,colback=white,arc=0mm,left={1mm},top={1mm},bottom={1mm},right={1mm},boxrule={0.25mm}]
\textit{\textbf{Summary of Answers to RQ2.1:}}
The analysis of the 5728 failed web-form-test submissions revealed three primary reasons for the failures: 
failure to generate web-form tests in the restricted format (40.82\%);
inability to generate correct test content based on the provided contextual information (29.50\%); and connection/API-related issues (29.68\%). 
\end{tcolorbox}

\subsubsection{Answer to RQ2.2}

\begin{figure*}[!t]
\centering
    \subfigure[RH-P]{\includegraphics[width=0.47\textwidth]{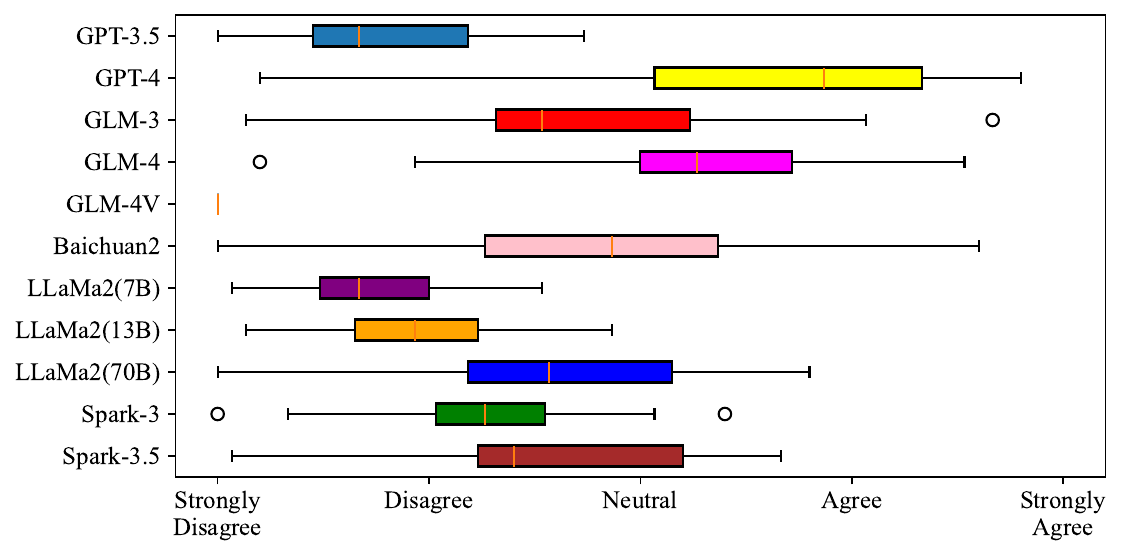}\label{FIG:RH-P quality}}\hspace{3mm}
    \subfigure[LH-P]{\includegraphics[width=0.47\textwidth]{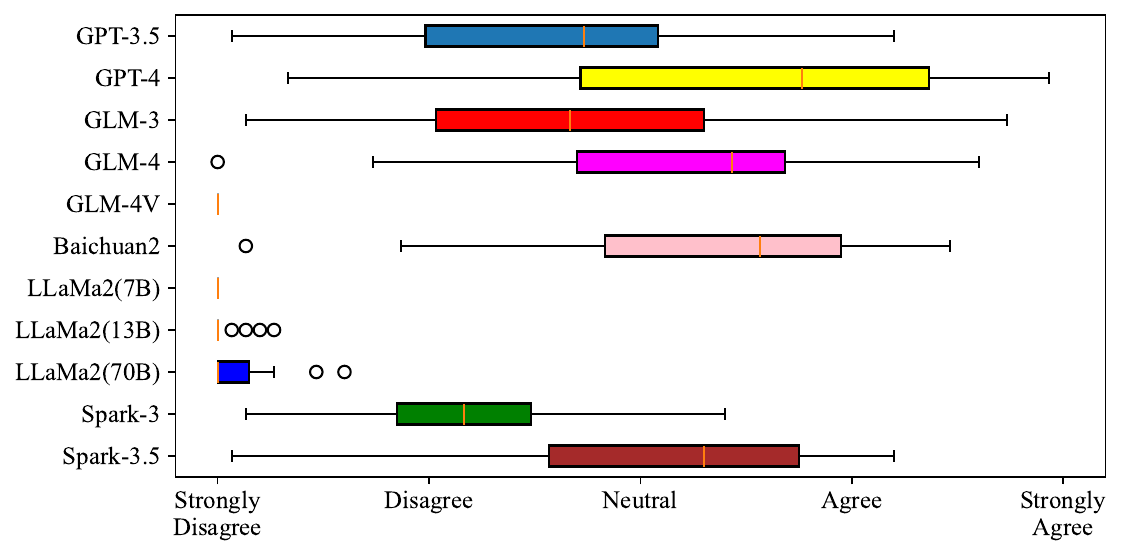}\label{FIG:LH-P quality}}

    \subfigure[PH-P]{\includegraphics[width=0.47\textwidth]{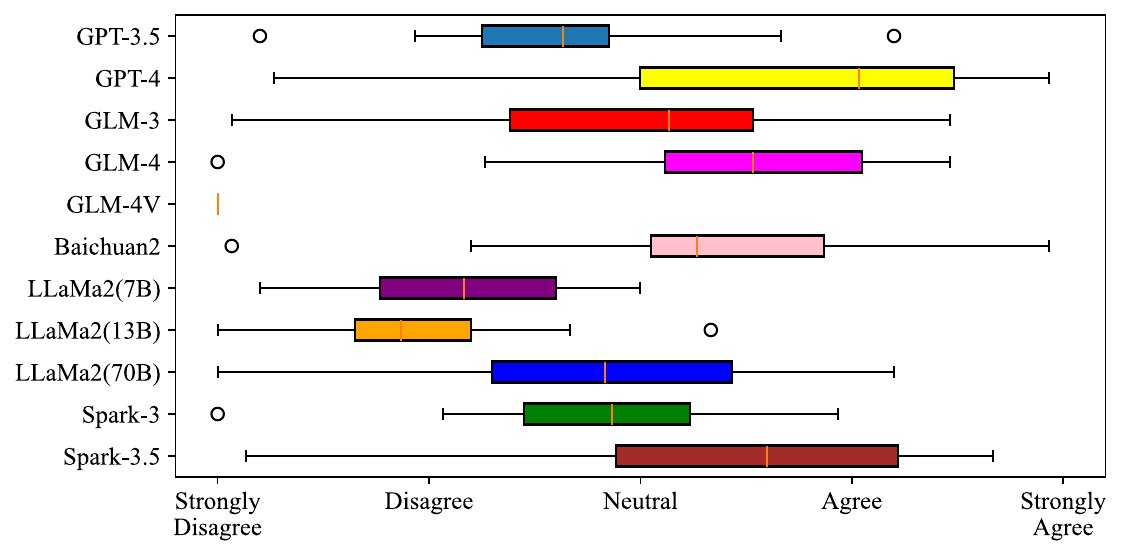}\label{FIG:PH-P quality}}\hspace{3mm}
    \subfigure[\textit{Average}]{\includegraphics[width=0.47\textwidth]{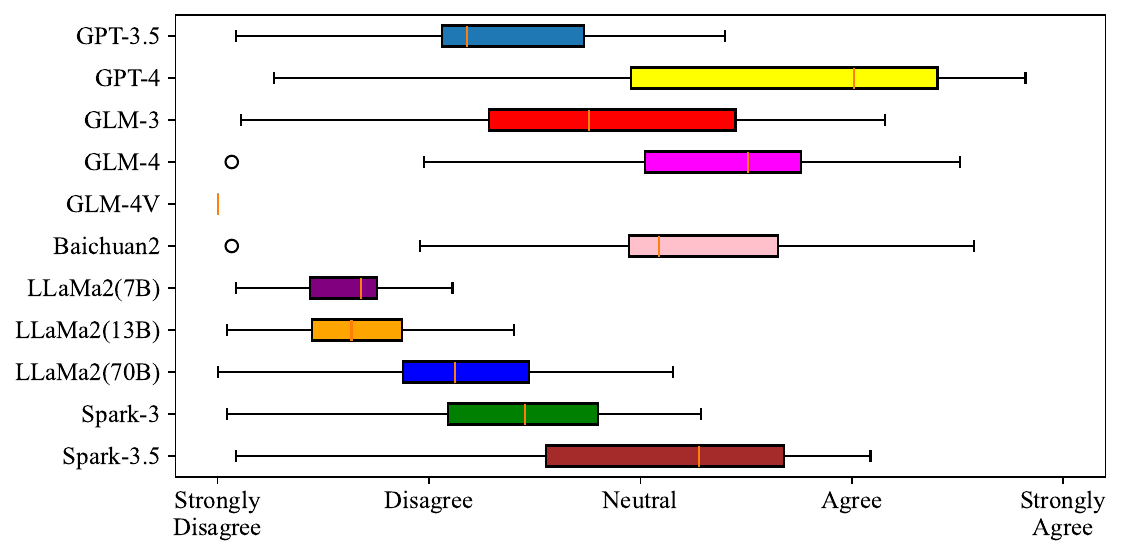}\label{FIG:average quality}}
\caption{Quality scores by 20 testers for different LLMs-generated web-form tests.}
\label{FIG: rq2 testers scores}
\end{figure*}

Figure \ref{FIG: rq2 testers scores} shows the quality scores given by 20 testers for the different LLM-generated web-form tests, using the three types of prompts.
Table \ref{TAB:rq2.2} presents the average quality scores of the 20 testers.
The testers were selected based on their expertise in software testing and development, with each holding at least a bachelor’s degree in computer science or a related field. 
On average, the testers had over seven years' experience in software testing, including experience in web-application testing and development, or academic research in relevant areas.
Based on the results, we have the following observations:

\begin{itemize}
    \item 
    The Kendall's W value~\cite{siegel1957nonparametric} for the 20 testers was $0.94$ which, being close to $1.0$, indicates a strong agreement among the evaluators.

    \item 
    The average scores for RH-P, LH-P, and PH-P were 2.40, 2.26, and 2.76, respectively.
    PH-P scored higher than the overall average (2.47), while LH-P and PH-P scored less.

    \item 
    For the RH-P prompt, only GPT-4 and GLM-4 had average scores greater than 3.00 (3.62 and 3.21, respectively), indicating a ``Neutral'' performance, according to the testers. 
    The testers rejected (sometimes strongly) the quality of the web-form tests generated by the other LLMs using RH-P.

    \item 
    For the LH-P prompt, GPT-4, GLM-4, Baichuan2, and Spark-3.5 were evaluated as having a ``Neutral'' performance, according to testers.
    The testers rejected (sometimes strongly) the quality of the web-form tests generated by the other LLMs using LH-P. 

    \item 
    For the PH-P prompt, GPT-4, GLM-3, GLM-4, Baichuan2, and Spark-3.5 were all evaluated as having a ``Neutral'' performance, according to the testers.
\end{itemize}

The average score across the 11 LLMs was 2.47. 
Five of the LLMs (GPT-4, GLM-3, GLM-4, Baichuan2, and Spark-3.5) scored better than the average.
GPT-4 always scored better than the other LLMs; and 
GLM-4V was always evaluated as the worst.

\begin{tcolorbox}[breakable,colframe=black,colback=white,arc=0mm,left={1mm},top={1mm},bottom={1mm},right={1mm},boxrule={0.25mm}]
\textit{\textbf{Summary of Answers to RQ2.2:}}
The main findings of the quality evaluation by the 20 testers are consistent with the findings of the effectiveness evaluation (Section \ref{sec: results-rq1}):
GPT-4 was the highest endorsed by the testers, with the other LLMs receiving poorer evaluations.
Overall, it appears that there is still room to improve the quality of the LLM-generated web-form tests, to better meet the expectations of the testers.
\end{tcolorbox}

\begin{table}[!t] 
\footnotesize
\setlength\tabcolsep{0.36mm} 
\caption{Average quality scores by 20 testers.} 
\label{TAB:rq2.2} 
\centering 
\begin{tabular}{@{}c|ccccccccccc|c@{}} 
\hline
\begin{tabular}[c]{@{}c@{}}\textbf{Prompt}  \\ \textbf{types}\end{tabular}&  \textbf{GPT-3.5} & \textbf{GPT-4} & \textbf{GLM-3} & \textbf{GLM-4} & \textbf{GLM-4V}  & \textbf{Baichuan2} & \begin{tabular}[c]{@{}c@{}}\textbf{LLaMa2}\\\textbf{(7B)}\end{tabular}  & \begin{tabular}[c]{@{}c@{}}\textbf{LLaMa2}\\\textbf{(13B)}\end{tabular} & \begin{tabular}[c]{@{}c@{}}\textbf{LLaMa2}\\\textbf{(70B)}\end{tabular} & \textbf{Spark-3} & \textbf{Spark-3.5} &\textbf{\textit{Average}}\\ \hline
RH-P & 1.82 & 3.62 & 2.76 & 3.21 & 1.00 & 2.84 & 1.70 & 1.98 & 2.63 & 2.26 & 2.60 & 2.40 \\ 
LH-P & 2.63 & 3.53 & 2.71 & 3.25 & 1.00 & 3.30 & 1.00 & 1.03 & 1.11 & 2.20 & 3.07 & 2.26 \\ 
PH-P & 2.63 & 3.71 & 3.01 & 3.45 & 1.00 & 3.38 & 2.16 & 1.95 & 2.81 & 2.81 & 3.43 & 2.76 \\ \hline
\textbf{\textit{Average}} & 2.36 & 3.62 & 2.83 & 3.31 & 1.00 & 3.17 & 1.62 & 1.65 & 2.18 & 2.42 & 3.03 & 2.47 \\\hline 
\end{tabular} 
\end{table}

\subsection{Insights for Using LLMs in Web-form Testing}

Based on the above findings, this section provides some insights for using LLMs to support the web-form-test generation, from the perspectives of \textit{prompt design} and \textit{LLM selection}.

\subsubsection{Insights for Designing Prompts}
Based on the experimental results for the three types of prompts (RH-P, LH-P, and PH-P), we found that the following strategies should be followed:

\begin{itemize}
    \item 
    \textbf{\textit{Insight 1: Ensure accurate extraction of web-form context for prompt guidance.}}
    We can extract the contextual information from the HTML of the web form and use it as part of the prompt to guide LLMs to generate web-form tests. 
    However, we need to accurately parse the web form and extract the contextual information. 
    For example, we found that using PH-P to construct prompts was 20.36\% more successful than using LH-P.
   
    \item 
    \textbf{\textit{Insight 2: Simplify the web-form HTML for simpler, clearer prompts.}}
    We found that the HTML structure of web forms can be complex.
    It is necessary to simplify this when constructing prompts, and to avoid directly using raw HTML to build prompts (which may make the content of the prompts too complex).
    For example, we found that using PH-P to construct prompts was 9.94\% more successful than using RH-P.
    
    \item 
    \textbf{\textit{Insight 3: Simple and clear task requirements should be set for LLMs to help them achieve the expected results more effectively.}}
    The PH-P method extracts web-form contextual information using pruned web-form HTML content, while 
    RH-P directly uses the pruning HTML (more complex) to construct contextual information:
    PH-P provides LLMs with simpler contextual information.
    We found that simpler and clearer prompts achieve more effective web-form-test generation.
\end{itemize}

\subsubsection{Insights for Selecting LLMs}

According to our study\textsuperscript{\ref{footnote:study}}, over 50\% (11 out of 20) of the testers have started using LLMs to guide quality-assurance work, which indicates that LLMs are playing an increasingly important role in testing.
80\% (16 out of 20) of the testers were concerned about privacy and security issues when using LLMs for the web-form quality assurance.
Based on the importance of LLMs, the security concerns of testers, and the main findings of our research, we offer the following guidance for selecting LLMs for testing:

\begin{itemize}
    \item 
    \textbf{\textit{Insight 4: To generate effective web-form tests, GPT-4 is highly recommended.}}
    According to our experimental results 
    (Section \ref{sec: results-rq1}), 
    GPT-4 can generate the most effective web-form tests.

    \item 
    \textbf{\textit{Insight 5: If facing testing constraints, alternative LLMs (not the GPT series) should be selected.}}
    Actual web-form testing tasks can involve combining real user data into prompt content. 
    This helps the LLMs better understand the contextual information of the web forms. 
    However, this may also lead to private data leakage:
    GPT LLMs should, therefore, not be directly used~\cite{wu2024new}.
    Alternative LLMs, such as the LLaMa2 series LLMs, should be considered.
    In this study, we compared the effectiveness of GPT-3.5 and GPT-4 with other LLMs, which should help testers more conveniently choose appropriate LLMs based on the actual situation.
    For example, the GLM-4 and Baichuan2 effectiveness are only about 10\% less than GPT-4.
\end{itemize}

\subsubsection{Insights for Testers}

Our findings provide concrete guidance on how testers can improve automated web-form-test generation using LLMs. 
To optimize test effectiveness, we recommend the following strategies:
\begin{itemize}
    \item 
    \textbf{\textit{Insight 6: Prioritize structured prompts for higher SSRs.}}
    Our results show that PH-P delivers better SSRs than RH-P and LH-P. 
    Testers should avoid raw HTML input and instead preprocess web-form structures to create well-structured prompts.

    \item 
    \textbf{\textit{Insight 7: Select high-performing LLMs for optimal results.}}
    GPT-4, GLM-4, and Baichuan2 are the most effective models for web-form testing. 
    If privacy constraints exist, alternative models like LLaMa2-70B should be considered 
    (though at the cost of lower SSRs).

    \item 
    \textbf{\textit{Insight 8: Ensure clear and concise test prompts.}}
    PH-P's superior performance suggests that reducing unnecessary HTML complexity improves LLM interpretation and response accuracy. 
    Testers should refine prompt structures by removing redundant form elements and explicitly defining expected input formats.
\end{itemize}

\subsection{Threats to Validity}

This section discusses some potential threats to the validity of our study.

The first threat relates to the representativeness of the experimental subject and baselines used in the study. 
To mitigate this threat, when selecting the AUTs on Github, we filtered through multiple keywords (such as  \textit{``Java Web''}, \textit{``Jobs''}, and \textit{``Books''}). 
This enabled the selection of different types of Java web applications.
We also conducted an in-depth investigation and analysis of current mainstream and widely-used LLMs, before making our selection.
Because our study focused on the testing effectiveness of various LLM-generated web-form tests, we did not include other baseline methods (such as random-based approaches~\cite{Alshraideh2006,Tonell2004}) in our comparison.
 
The second threat involves our research on web-form contextual information.
We designed three types of prompts to better extract contextual information from the web forms.
This helped us to build multiple prompts to guide the LLMs' generation of web-form tests, and to evaluate their effectiveness.
Following previous research~\cite{liu2023fill}, to reduce the randomness and non-determinism~\cite{Ouyang23} in the LLM responses, we also conducted the experiments for each type of prompt three times.

The third threat refers to whether or not the studies were conducted fairly.
In this study, we used the successfully-submitted rates (SSRs) as the metric to evaluate the testing effectiveness of different LLM-generated web-form tests.
Another popular metric for evaluation~\cite{Mesbah2012,Li2014} is \textit{code coverage}, which refers to the extent to which the source code is executed during testing. 
A complete code coverage analysis requires testing the entire application, including various interconnected components and flows.
In our study, because we targeted specific components that include web forms, rather than evaluating the entire web application, code coverage was not a suitable metric.

The fourth threat relates to the applicability of our approach to closed-source (proprietary) projects. 
Since the training data of LLMs is predominantly derived from open-source repositories~\cite{chang2023survey, naveed2023comprehensive}, the generated web-form tests may reflect the characteristics and conventions of open-source development. 
In our study, web forms were obtained by parsing HTML content from publicly accessible URLs, which primarily originate from open-source applications. 
Closed-source projects may employ proprietary frameworks or custom HTML structures, which may introduce variations in form parsing and data handling, potentially affecting the ability of LLMs to extract and interact with their web forms.
While our methods were designed based on the general properties of web forms, such as input fields and submission logic, we acknowledge that the absence of closed-source applications in our experiments limits the generalizability of our findings. 
Nevertheless, we believe that, fundamentally, our approach retains applicability across a wide range of scenarios.

The fifth threat relates to the uniformity of prompt design across different LLMs. While we employed three types of prompts (RH-P, LH-P, and PH-P) to extract web-form contextual information, our results show that different LLMs respond to the same prompts with varying effectiveness.
While GPT-4 and Baichuan2 achieved high SSRs, models like LLaMa2(7B/13B) and GLM-4V performed significantly worse, suggesting that a standardized prompt structure may not be equally effective across all models.
Additionally, there may be other potential prompt-construction methods that could make the web-form extraction process more efficient.
Future work may explore these methods and assess their applicability to improve the web-form testing process.

The final threat relates to the validation and validity of LLM-generated web-form tests. 
When testing a system, it is essential to consider both \textit{valid} and \textit{invalid} tests: 
The former typically guides normal execution and validates the program's functionality; 
while the latter is used to test the program's ability to detect and handle exceptions~\cite{shahbaz2015automatic}.
In this study, we focused on evaluating the correctness of submission statements when a web form receives a set of valid tests: 
If the AUT returned a successful response, then the submission was considered successful.
However, some invalid data (such as ``user@'' or ``@email.com'' 
---
incomplete email addresses) may also be successfully submitted due to a lack of exception-handling processes.
The validation of such invalid tests will form part of our future work.

\section{Related Work
\label{SEC:Related Work}}

This section examines four areas of related work: 
web-form testing, GUI Testing, string-data generation, and LLMs for test-case generation.

\subsection{Web-Form Testing}
Web-form testing is critical to software quality assurance, impacting both the user experience and the application's stability. 
As web applications become increasingly complex, researchers have worked to enhance automation, efficiency, and accuracy in web-form testing. 

Rothermel et al.~\cite{rothermel1998you} proposed an automated framework for identifying and validating form elements and their interaction behavior. 
Using the consistency between visual elements and back-end logic, their method significantly improved testing coverage and efficiency, setting new standards for form-based visual-program testing.
Ricca et al.~\cite{ricca2001analysis} introduced a UML model to assess static site structures and guide white-box testing. 
Applied to real-world scenarios, their approach improved verification and validation, using automatic test-case generation to ensure comprehensive testing and to simplify regression checks. 
They emphasized the importance of thorough testing to ensure the quality and performance of web applications, offering a detailed perspective on web-form testing.
Furche et al.~\cite{furche2013ontological} used ontology tools to automate the understanding of form structure and content: 
They optimized the integration and retrieval process of form data, and demonstrated the efficiency and potential advantages of semantic technology for handling complex web forms.
Santiago et al.~\cite{santiago2019machine} integrated machine learning and constraint-solving techniques to predict the effective input of form fields.
They were able to generate test cases that comply with logical constraints, demonstrating the practical application of machine learning technology in enhancing test automation and improving accuracy.
Cruz-Benito et al.~\cite{cruz2017enabling} explored the adaptability of web forms based on user-feature detection. 
They studied the impact of user behavior and preferences on form design through A/B testing and machine learning techniques, and proposed strategies to improve user satisfaction and interaction efficiency through customized experiences.
Lukanov et al.~\cite{lukanov2016using} used the \textit{functional Near Infrared Spectroscopy} (fNIRS) measure to study the impact of web-form layout on users' psychological burden. 
This provided the scientific basis for understanding how different design schemes affect users' cognitive burden, emphasizing the importance of optimizing the user experience in form design.
Alian et al.~\cite{alian2024bridging} improved the accuracy and efficiency of automatic filling and validation processes by conducting an in-depth analysis of the semantics of form elements.
They highlighted the critical role of semantic analysis in improving the quality of web-form testing.

\subsection{GUI Testing}

Graphical User Interface (GUI) testing is critical to ensuring the quality and usability of modern software applications. 
As web applications become increasingly complex and dynamic, researchers have developed various approaches to enhance the accuracy, efficiency, and scalability of GUI testing. 
These approaches address challenges such as cross-browser consistency, fault detection, and robust test-case generation.

Xu et al.~\cite{xu2018cross} proposed an empirical metric for detecting cross-browser visual differences in web pages. 
Their approach translated the Gestalt laws of grouping into computational rules, utilizing a block tree structure to evaluate proximity, color, and image similarity. 
Experiments on widely-used web pages validated its applicability for addressing rendering consistency issues.
Moura et al.~\cite{moura2023cytestion} proposed Cytestion, a scriptless tool for automated GUI testing in web applications. 
The tool evaluates HTTP request statuses, failure messages, and console logs to detect faults such as crashes and errors. Its effectiveness was assessed through empirical studies on open-source and industrial applications.
De Almeida Neves et al.~\cite{de2022morpheus} introduced Morpheus Web Testing, a tool that automates functional test-case generation for widget-based web applications by extracting widget information from Java Server Faces (JSF) artifacts. 
A case study comparing Morpheus with CrawlJax highlighted its ability to achieve higher code coverage and improve automated testing processes.
Nguyen et al.~\cite{nguyen2021generating} introduced a semantic-based method for generating and selecting resilient locators in web UI testing. 
Using neighboring elements to construct XPaths, the method addresses locator instability and was evaluated on 15 websites to assess accuracy and maintainability.
De Luca et al.~\cite{de2024investigating} investigated a hook-based approach for improving locator robustness in template-based web-application testing. 
Hooks, injected as HTML tag attributes, uniquely identify GUI elements to reduce locator fragility.
A three-dimensional model for classifying layout changes enabled systematic analysis of locator robustness across test-generation techniques.
Qi et al.~\cite{qi2019leveraging} proposed a keyword-guided exploration strategy for constructing web-application test models. 
By calculating similarity scores between keywords and web-page content, the method prioritizes relevant UI states and transitions. 
It was evaluated on nine web applications, showing its ability to achieve functionality coverage within a time budget.
Mattiello et al.~\cite{mattiello2022model} proposed a model-based testing approach that infers models from existing automated tests to generate new test cases. 
Using event-driven models with parameters and input data, the method was applied to extend system-level GUI tests in web applications.
Wang et al.~\cite{wang2024leveraging} introduced VETL, a web-testing technique driven by
large vision-language models (LVLMs). 
VETL generates text inputs using the scene-understanding capabilities of LVLMs and formulates GUI-element selection as a visual question-answering task. 
It employs a multi-armed bandit module to guide exploration in dynamic GUI contexts.
Fan et al.~\cite{fan2023comprehensive} proposed a Q-learning-based framework for automatic web-GUI testing. 
Four QL-specific configurations were systematically evaluated on two open-source web applications and one industrial website, standardizing experimental settings and analyzing QL's role in exploring application states.

\subsection{String-Data Generation}

String-data generation is essential in software testing for validating functionality and detecting defects. 
Various methods have been proposed to improve the diversity and effectiveness of generated string inputs, including random-based~\cite{miller1990empirical}, search-based~\cite{alshraideh2006search,mcminn2012search}, and model-driven approaches~\cite{almeida2007enumeration,wang2013probabilistic}.

Miller et al.~\cite{miller1990empirical} introduced fuzz testing, which uses random character generators to create unpredictable inputs, effectively exposing software vulnerabilities and crashes. This approach proved effective in identifying robustness issues in UNIX utilities.
Alshraideh et al.~\cite{alshraideh2006search} proposed a technique that uses program-specific search operators, customizing strategies based on program structure and logic to improve the relevance and coverage of test data.
Almeida et al.~\cite{almeida2007enumeration} used string automata to create an automaton model to generate all possible strings under specific conditions:
This provided an improved understanding of program behavior with various string inputs, and enhanced test comprehensiveness. 
Zhao et al.~\cite{zhao2010automatic} introduced an automatic string-data generation method to efficiently identify domain errors and boundary conditions causing abnormal program behaviors.
McMinn et al.~\cite{mcminn2012search} proposed a method of generating test data by combining the results of web searches, and analyzing the results of network queries to obtain string data from actual application scenarios, thus producing test inputs that are more closely related to real-world situations.
Wang et al.~\cite{wang2013probabilistic} developed a probabilistic model for predicting and generating the results of string transformations, offering a new perspective on understanding how strings change in specific applications by analyzing the characteristics and transformation patterns of the string data.
Shahbaz et al.~\cite{shahbaz2015automatic} streamlined test-data creation for string validation by using web searches and regular expressions, efficiently producing various valid and invalid inputs to test string logic.
Liu et al.~\cite{liu2023fill} investigated an automated text-input generation method for mobile GUI testing:
It considered the context information of the application to automatically produce text-input tests, testing the responsiveness and correctness of the GUI.

\subsection{LLMs for Test-Case Generation}

LLMs have been increasingly applied to automate test-case generation, demonstrating their ability to produce diverse and context-aware test cases. 
Recent studies have explored their applications in unit testing, security testing, and flaky-test prediction.

Yu et al.~\cite{yu2023llm} explored the use of LLMs for improving the generation and migration of automated test scripts, noting their flexibility and efficiency in complex scenarios, and the potential for enhanced script maintainability and adaptability.
Schäfer et al.~\cite{schafer2023empirical} conducted an empirical evaluation of the effectiveness of LLMs for automated unit-test generation, highlighting the potential for LLMs to generate high-quality test cases while also noting their limitations in dealing with specific testing challenges.
Zhang et al.~\cite{zhang2023well} evaluated the effectiveness of LLM-generated security tests. 
Through experimental comparisons, they demonstrated LLMs' ability to identify potential security vulnerabilities and generate corresponding test cases, highlighting their potential for automating security testing.
Fatima et al.~\cite{fatima2022flakify} proposed a black-box approach based on LLMs for predicting flaky tests. 
By analyzing the historical test-execution data, this predictor can effectively identify potentially flaky tests, helping to improve the stability and reliability of testing.

To the best of our knowledge, there has been no empirical research assessing the effectiveness of LLMs in generating web-form tests. 
Additionally, there is limited understanding of how testers can make optimal use of LLMs for this purpose. 
This article aims to address these gaps in the literature.

\section{Conclusions and Future Work
\label{SEC:Conclusions and Future Work}}

This paper has reported on an empirical study to investigate the effectiveness of 11 state-of-the-art LLMs in web-form-test generation. 
We evaluated these LLMs using 146 web forms from 30 open-source Java web applications.
Based on our experimental results, we have the following conclusions:
(1) Some LLMs (such as GPT-4, GLM-4, and Baichuan2) can generate relatively efficient and high-quality tests for the web form. 
However, under the same conditions, LLMs such as GLM-4V, LLaMa2(7B), and LLaMa2(13B) did not perform well on the same testing task, indicating that LLMs can still be optimized and improved for automated web-form-test generation.
(2) Some LLMs (such as GLM-3, GLM-4, Baichuan2, and Spark-3.5) may be more suitable for generating appropriate tests for web forms than GPT-3.5, delivering a higher SSR than GPT-3.5.
(3) A comparison of the experimental results for the three different prompt methods (RH-P, LH-P, and PH-P) revealed that clear and concise web-form contextual content could better guide the LLMs to generate appropriate content.
If the contextual information of key information in the HTML elements is missing, it may reduce the effectiveness by around 10\% to 20\%.
(4) Regardless of the prompt, GLM-4V performs poorly. 
An analysis of this found that GLM-4V may not be suitable for generating web-form tests,  which indicates that the selection of models should depend on their specific areas of expertise.

Our future research will include the following directions:
\begin{itemize}
    \item 
    Based on the experimental results, the average SSR of the selected LLMs was 60.37\%, which means that there is room for improvement.
    Some approaches can be adopted to improve the LLM's effectiveness for generating web-form tests, such as optimizing prompt design, fine-tuning LLMs, and adapting prompts to different models.
    Future work could investigate automated prompt-optimization techniques, such as reinforcement learning, to refine prompt structures based on model-specific performance.
    
    \item 
    An automated web-form-test generation tool with various LLM options shall be designed and developed.
    This tool will be able to provide testers with intelligently recommended options (guided by different LLMs) for the testing process, and more comprehensive testing analysis from the perspective of code coverage, bug detection, etc.
    Future work could expand the tool to support other web application types, such as dynamic content or multi-step workflows, and it could also be extended to mobile and desktop applications with different GUI structures.
    
    \item 
    We found that some web forms may lack essential exception-handling processes, which could allow invalid web-form tests to be successfully submitted. 
    This could lead to security issues, such as XSS or SQL injection~\cite{goto2022Design}. 
    Therefore, exploring how to generate effective invalid tests to detect these vulnerabilities is an important direction for future research.
    This could involve extending the tests to APIs and dynamic web pages to assess the robustness of the system in diverse scenarios.

    \item 
    Since many web forms require the submission of personal data (such as names, email addresses, and passwords), the risk of data leakage is a significant concern~\cite{yao2024survey}.
    Exploring the capability of LLM-generated web-form tests to detect vulnerabilities related to data privacy will also form part of our future work.
\end{itemize}

\section*{Acknowledgment}
The authors would like to thank the anonymous reviewers for their many constructive comments.
We would like to thank the testers who participated in our study, as well as Yujie Xie and Yinming Huang from the Macau University of Science and Technology for their valuable help in the experiments.
This work is partly supported by the Science and Technology Development Fund of Macau, Macau SAR, under Grant No. 0021/2023/RIA1.

{
\bibliographystyle{ACM-Reference-Format}
\bibliography{ref.bib}
}

\end{document}